\documentclass[%
 reprint,
 amsmath,amssymb,
 aps,
]{revtex4-2}

\usepackage{graphicx}
\usepackage{dcolumn}
\usepackage{bm}
\usepackage{float}
\usepackage{mleftright}
\usepackage{comment}
\usepackage{amsmath}
\usepackage{multirow}
\usepackage[american]{circuitikz}
\usetikzlibrary{calc}
\usepackage{physics}
\usepackage{siunitx}
\usepackage[normalem]{ulem}
\usepackage[colorlinks]{hyperref}
\hypersetup{%
        plainpages=true,
        breaklinks=true,
        hypertexnames=false,
        pageanchor=true,
        colorlinks=true,
        linkcolor={blue},
        citecolor={blue},
        urlcolor={blue},
        anchorcolor={black}
      }

\usepackage{mleftright} 

\newcommand{\figref}[1]{\mbox{Fig.~\ref{#1}}}
\newcommand{\tabref}[1]{\mbox{Table~\ref{#1}}}
\newcommand{\secref}[1]{\mbox{Sec.~\ref{#1}}}

\newcommand{\appref}[1]{\mbox{Appendix~\ref{#1}}}
\renewcommand{\eqref}[1]{\mbox{Eq.~(\ref{#1})}}

\newcommand{\figpanel}[2]{Fig.~\hyperref[#1]{\ref*{#1}(#2)}}
\newcommand{\figpanels}[3]{Fig.~\hyperref[#1]{\ref*{#1}(#2)--(#3)}}
\newcommand{\figpanelNoPrefix}[2]{\hyperref[#1]{\ref*{#1}(#2)}}

\allowdisplaybreaks

\begin{document}

\preprint{APS/123-QED}

\title{Suppressing microwave crosstalk in superconducting quantum processors \\ by designing parasitic drive capacitances}

\author{Anuj Aggarwal}
\author{Jorge Fern\'andez-Pend\'as}
\author{Simon Pettersson Fors}
\author{Zakaria Mohamed}
\author{Halld\'{o}r Jakobsson}
\author{Amr Osman}
\author{Janka Bizn\'{a}rov\'{a}}
\author{Andreas Nylander}
\author{Anita Fadavi Roudsari}
\author{Marcus Rommel}
\author{Anton Frisk Kockum}
\author{Giovanna Tancredi}
  \email{tancredi@chalmers.se}

\affiliation{Department of Microtechnology and Nanoscience, 
Chalmers University of Technology, SE-412 96 Gothenburg, Sweden}

\date{\today}


\begin{abstract}
As superconducting quantum processors scale to large qubit counts, unwanted microwave drive crosstalk becomes an important source of coherent errors. Here, we investigate the role of parasitic coupling paths in the suppression of drive crosstalk in a processor with two fixed-frequency transmon qubits and a tunable coupler. To account for the many different coupling paths, we develop a detailed circuit model to capture the dependence of crosstalk on the coupler frequency. We experimentally demonstrate suppression of crosstalk, finding that the suppression does not originate solely from cancellation of the effective qubit-qubit coupling, but also is a result of interference between other coupling paths. Building on these insights, we show that drive crosstalk can be suppressed with the right choice of parasitic capacitances, which otherwise often are treated as unwanted design imperfections. Our results provide practical guidance for low-crosstalk device design of superconducting quantum processors and point toward a broader strategy in which parasitic coupling paths are engineered rather than simply minimized.

\end{abstract}

\maketitle


\section{Introduction}

With superconducting quantum processors now reaching the scale of hundreds of qubits~\cite{google2025quantum, kim2023evidence, gao2025establishing}, precise control of individual qubits becomes increasingly challenging~\cite{Mohseni2026}. Microwave drives intended to control the quantum state of a qubit (target) can affect other qubits (victims) in the processor, an effect known as drive crosstalk.  A near-resonant signal can directly drive transitions in the victim qubits, while an off-resonant signal can induce an AC Stark shift and cause unwanted phase accumulation~\cite{ketterer2023characterizing, nuerbolati2022canceling}. Both effects lead to coherent errors and reduce the fidelity of quantum operations. Although pulse-compensation techniques can mitigate these effects~\cite{wesdorp2026mitigating, nuerbolati2022canceling, sung2021realization, Huang2026}, such techniques become increasingly demanding to implement and maintain as the number of qubits grows. 

Drive crosstalk originates from parasitic electromagnetic coupling between the qubits' control lines and other circuit elements in the quantum processor~\cite{kosen2024signal, spring2022high, song2026microwave, huang2021microwave, wenner2011wirebond}. For this reason, crosstalk is highly device-specific and does not necessarily follow the physical proximity of the qubits. For example, crosstalk can be mediated by the printed circuit board (PCB) hosting the processor~\cite{kosen2024signal, huang2021microwave}, electromagnetic box-modes in the sampleholder~\cite{huang2021microwave, song2026microwave}, and even wirebonds~\cite{wenner2011wirebond}. Generally, on-chip parasitic coupling paths are also regarded as sources of unwanted interactions. However, such coupling paths can also interfere with one another, allowing partial or even near-complete crosstalk cancellation at specific operating conditions. Understanding how these coupling mechanisms combine is therefore important both for characterizing crosstalk and, more practically, for identifying ways to suppress it at the device level.

In this article, we investigate the role of on-chip parasitic coupling paths in suppressing drive crosstalk. We develop a detailed theoretical model, taking into account many circuit elements, to characterize crosstalk as a function of the relevant circuit parameters in a processor containing two fixed-frequency superconducting transmon qubits coupled by a tunable coupler. We demonstrate, both theoretically and experimentally, that one can suppress crosstalk by tuning the drive-line coupling paths. Our results demonstrate that parasitic capacitances, which are often neglected during the design of superconducting quantum processors, influence crosstalk and can be exploited to mitigate it. More broadly, our insights highlight the importance of accounting for unintended coupling pathways in quantum processor design.

The article is organized as follows. In \secref{sec:model}, we derive an effective Hamiltonian model for our device, starting from a detailed circuit model, and obtain analytical expressions for crosstalk. Then, in \secref{sec:Experiment}, we report crosstalk experiment results on our device and compare them with our theoretical model, finding excellent agreement. In \secref{sec:Discussion}, we discuss the implications of our theoretical and experimental findings and present practical design guidelines for superconducting quantum processors. We summarize our conclusions in \secref{sec:Conclusion}. For the interested reader, we provide additional details about the theoretical model and the experiment in the appendices.


\section{Model}
\label{sec:model}

\begin{figure}
    \centering
    \includegraphics[width=\columnwidth]{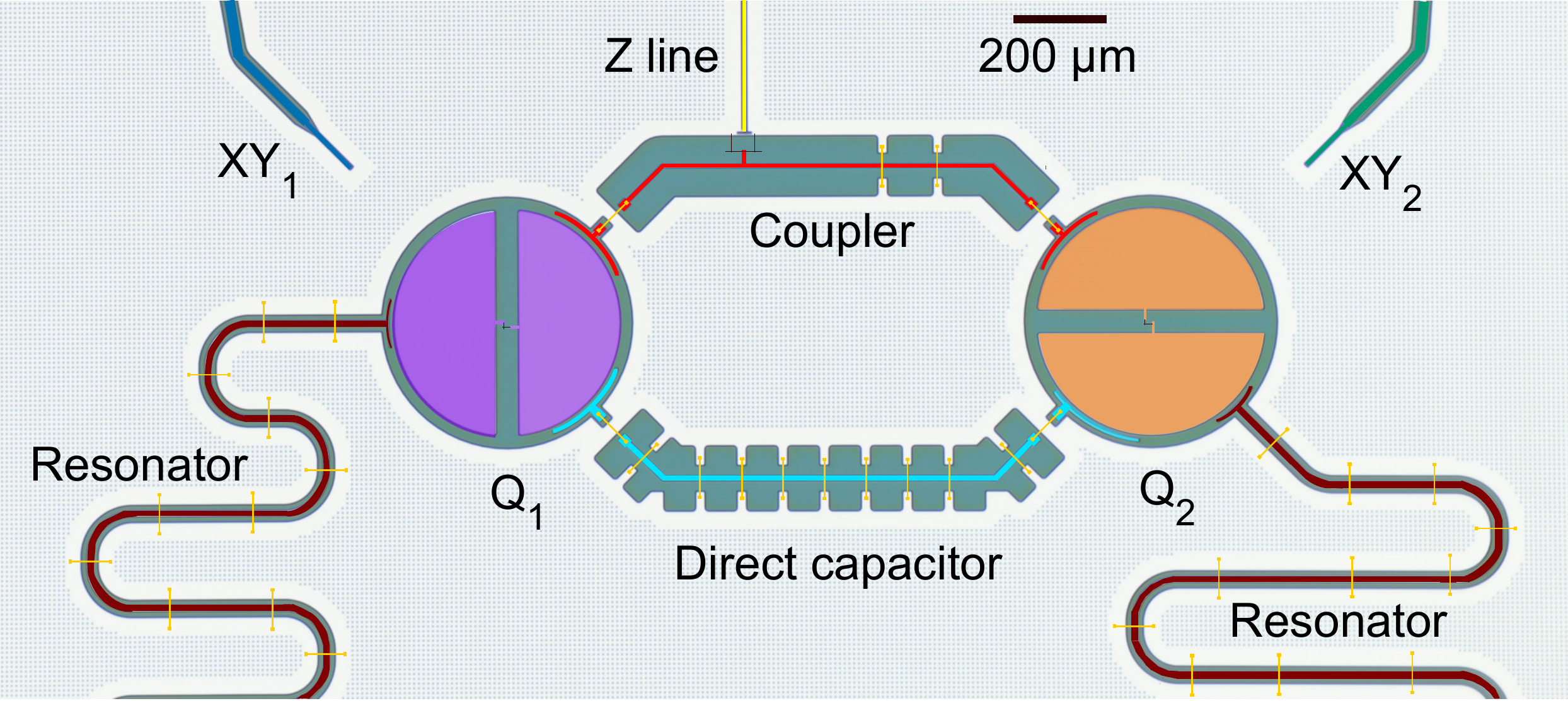}
    \caption{Device micrograph with false coloring. Two floating qubits ($\rm Q_{1}$ and $\rm Q_{2}$) are coupled via a tunable coupler and a direct capacitor. Each qubit has its own XY control line and readout resonator. The full list of the device parameters is given in \tabref{tab:floating_system_params} in \appref{app:exp_setup}.}
    \label{fig:device}
\end{figure}

\begin{figure*}
    \centering
    \newcommand{\schematicwidth}{0.8\textwidth}
    \input{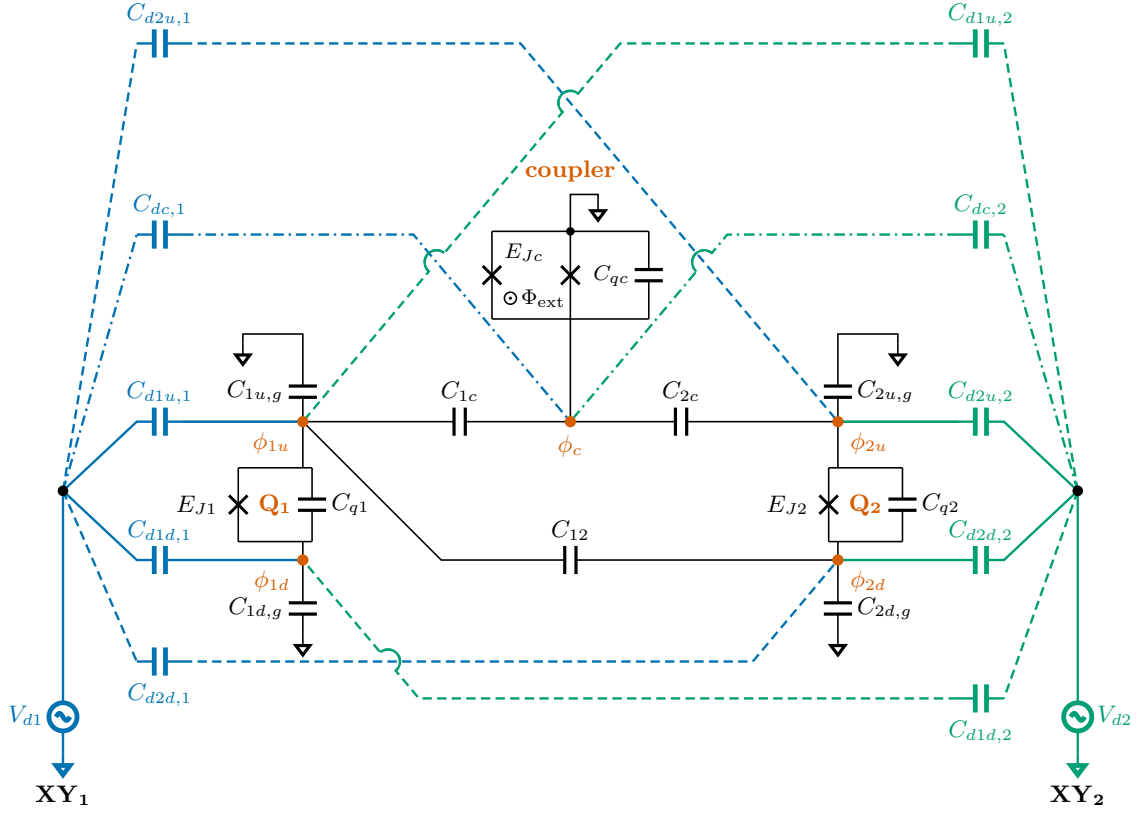}
\caption{Circuit schematic of two floating transmon qubits, Q$_1$ and Q$_2$, capacitively coupled to a grounded flux-tunable coupler through $C_{1c}$ and $C_{2c}$, with direct qubit--qubit coupling through $C_{12}$. The node fluxes $\phi_{1u}$, $\phi_{1d}$, $\phi_{2u}$, $\phi_{2d}$, and $\phi_c$ define the five circuit nodes, where `u' and `d' label the up and down flux nodes of each qubit. The XY$_1$ (blue) and XY$_2$ (green) drive lines couple capacitively to the circuit. Solid lines indicate coupling to the target qubit, dash-dotted lines coupling to the coupler, and dashed lines coupling to the victim qubit. The corresponding drive capacitances are labeled in the schematic. We neglect the small parasitic capacitive couplings between $\phi_{1u}-\phi_{2u}$, $\phi_{1d}-\phi_{2d}$, and  $\phi_{1d}-\phi_{2u}$, since they do not significantly change the effective charging energies of the system.}
    \label{fig:circuit_schematic}
\end{figure*}

To understand crosstalk mechanisms, we first derive an effective Hamiltonian model for our device. Starting from a circuit Lagrangian, we construct the full Hamiltonian. We then perform a high-frequency expansion to arrive at an effective time-independent Hamiltonian, which enables us to extract analytical expressions for the crosstalk.

The device consists of two fixed-frequency floating superconducting transmon qubits~\cite{koch2007charge}, Q$_1$ and Q$_2$, and a flux-tunable grounded transmon acting as a coupler, along with XY drive lines for each qubit (XY$_1$ and XY$_2$), as shown in \figref{fig:device}. In the circuit model of the device, shown in \figref{fig:circuit_schematic}, the most prominent capacitive couplings are those between each qubit and the coupler, and the ones between each drive line and the pads of its target qubit. However, we also include the direct qubit-qubit coupling, the parasitic capacitive coupling of the drive lines to the coupler, and the parasitic capacitive coupling of each drive line to the victim qubit. All these parasitic couplings turn out to be important for understanding the origins of crosstalk and for engineering its suppression.


\subsection{Circuit Lagrangian and bare Hamiltonian}
\label{sec:CircuitLagrangian}

The Lagrangian of the circuit in \figref{fig:circuit_schematic} can be written in the basis of the node fluxes $\phi^T = (\phi_{1u} , \phi_{1d} , \phi_{2u} , \phi_{2d}, \phi_{c})$ as
\begin{align}
  \mathcal{L}
  =&\ \frac{1}{2} \dot{\phi}^T C_m^{(0)} \dot{\phi} - \dot{\phi}^T q_d^{(0)} + E_{J1} \cos \mleft( \frac{\phi_{1u} - \phi_{1d}}{\varphi_0} \mright) \nonumber \\
  &+ E_{J2} \cos \mleft( \frac{\phi_{2u} - \phi_{2d}}{\varphi_0} \mright) + E_{Jc} (\Phi_{\text{ext}}) \cos \mleft( \frac{\phi_c}{\varphi_0} \mright) ,
  \label{eq:Lagrangian}
\end{align}
where $C_m^{(0)}$ is the capacitance matrix of the network (see \appref{app:LagrangianToHamiltonian}). Here, $E_{J1}$, $E_{J2}$, and $E_{Jc}$ are the Josephson energies of the qubits ($Q_{1}$ and $Q_{2}$) and the coupler, respectively. The coupler has a symmetric superconducting quantum interference device (SQUID), such that $E_{Jc}(\Phi_{\text{ext}})=2E_{Jc,0} \cos (\pi \Phi_{\text{ext}}/\Phi_0)$, where $E_{Jc,0}$ is the Josephson energy of each junction in the SQUID, $\Phi_{0}$ is the flux quantum, and $\varphi_{0} = \frac{\Phi_0}{2 \pi}$ is the reduced flux quantum. The charge-drive vector is
\begin{equation}
  q_d^{(0)} = V_{d1}
  \begin{pmatrix}
    C_{d1u,1} \\
    C_{d1d,1} \\
    C_{d2u,1} \\
    C_{d2d,1} \\
    C_{dc,1}
  \end{pmatrix} + V_{d2}
  \begin{pmatrix}
    C_{d1u,2} \\
    C_{d1d,2} \\
    C_{d2u,2} \\
    C_{d2d,2} \\
    C_{dc,2}
  \end{pmatrix}
\end{equation}
where `u' and `d' label the up and down flux nodes of each qubit in the schematic, and all other notation is defined in \figref{fig:circuit_schematic}. 

Going from the Lagrangian in \eqref{eq:Lagrangian} to a Hamiltonian describing the system requires several steps, which we describe in detail in \appref{app:LagrangianToHamiltonian}. Briefly put, we first rewrite the Lagrangian in terms of the qubits' common-mode ($+$) and differential-mode ($-$) fluxes:
\begin{align}
  \phi_{1\pm} =&\ \phi_{1u} \pm \phi_{1d} \, , \nonumber \\
  \phi_{2\pm} =&\ \phi_{2u} \pm \phi_{2d} \, .
\end{align}
The common-mode fluxes of the qubits are cyclic coordinates in the Lagrangian, and therefore the associated conjugate variables are conserved quantities. This allows us to ignore them and focus only on the differential modes of the qubits and on the coupler node flux. After a Legendre transform restricted to the differential modes, we obtain an effective Hamiltonian.

Finally, by quantizing this effective Hamiltonian, the system can be understood as three coupled weakly anharmonic (Kerr) oscillators with annihilation (creation) operators $a_k$ ($a_k^\dag$) for $k \in \{1,2,c\}$. Introducing the notation $X_k \equiv a_k^\dag - a_k$, we obtain the Hamiltonian
\begin{align}
  \mathcal{H} =&\ \sum_{k \in \{1,2,c\}} \omega_k a_k^\dag a_k \nonumber \\
  &+ \sum_{k \in \{1,2,c\}} \frac{\alpha_k}{2}\mleft[ a_k^{\dag 2} a_k^2
  + \mleft( a_k^{\dag 2} + \tfrac{2}{3} a_k^{\dag 3} a_k + \tfrac{1}{6} a_k^{\dag 4} + \text{H.c.} \mright) \mright] \nonumber \\
  &+ \sum_{k \in \{1,2,c\}} i \epsilon_k X_k
  + \sum_{ij \in \{1c,2c,12\}} g_{ij} X_i X_j \, . 
  \label{eq:kerr_h}
\end{align}

All parameters of the bare Hamiltonian in \eqref{eq:kerr_h} can be expressed in terms of the circuit parameters in \figref{fig:circuit_schematic}: the Kerr-oscillator frequencies of the qubits and the coupler are approximately given by~\cite{koch2007charge, fors2024comprehensive}
\begin{equation}
  \omega_{k} \approx \sqrt{8 E_{c,kk} E_{Jk}} - E_{c,kk} \, , \\
\end{equation}
their anharmonicities are
\begin{equation}
  \alpha_{k} \approx - E_{c,kk} \, , \\
\end{equation}
the couplings between pairs of the three elements are
\begin{equation}
  g_{ij} \approx - E_{c,ij} \sqrt[4]{\frac{4 E_{Ji} E_{Jj}}{E_{c,ii} E_{c,jj}}} \, ,\,  \qquad ij \in \{12, 1c, 2c\} \, ,
\end{equation}
and the drive on each element is
\begin{align}
  \epsilon_k =&\ \Big[\frac{V_{d1}}{e}\mleft(E_{c,1k}\tilde{C}_{d1} + E_{c,2k}\tilde{C}_{d2} + E_{c,kc}\tilde{C}_{dc}\mright) \nonumber \\
  &+ \frac{V_{d2}}{e}\mleft(E_{c,1k}\breve{C}_{d1} + E_{c,2k}\breve{C}_{d2} + E_{c,kc}\breve{C}_{dc}\mright)\Big] \sqrt[4]{\frac{8 E_{Jk}}{E_{c,kk}}} \, .
\label{eq:DriveStrengths}
\end{align}

Here, the charging energy $E_{c,ij} = \frac{e^2}{2} \mleft[C_m^{-1}\mright]_{ij}$ is given by the inverse of the capacitance matrix transformed to the basis of common- and differential-mode fluxes. $\tilde C$ and $\breve C$ are the effective drive capacitances for $\rm{XY}_1$ and $\rm{XY}_2$, respectively, described by \eqref{eq:effective_drive_caps} in \appref{app:LagrangianToHamiltonian}. 
Since $E_{Jc}$ is flux-tunable, $g_{1c}$, $g_{2c}$, and $\epsilon_c$ all inherit a dependence on the external flux $\Phi_{\rm ext}$ as well.

From \eqref{eq:DriveStrengths}, we obtain the relative drive strengths of the victim qubit and the coupler with respect to the target qubit, considering one drive line at a time ($V_{d2}=0$ for XY$_1$ and $V_{d1}=0$ for XY$_2$). These ratios are
\begin{equation}
  \lambda_{ ji}
  = \frac{\epsilon_j}{\epsilon_i}
  = \frac{E_{c,1j} \tilde{C}_{d1} + E_{c,2j} \tilde{C}_{d2} + E_{c,jc} \tilde{C}_{dc}}{E_{c,1i} \tilde{C}_{d1} + E_{c,i2} \tilde{C}_{d2} + E_{c,ic} \tilde{C}_{dc}} \sqrt[4]{\frac{E_{Jj} E_{c,ii}}{E_{Ji} E_{c,jj}}} \, ,
\end{equation}
where $i \in \{ 1, 2 \}$ and $j \in \{ 1, 2, c \} \neq i$. However, these ratios only characterize the direct, capacitive coupling strength to each port. They are not yet the full crosstalk, since $g_{12}$, $g_{1c}$, and $g_{2c}$ open an additional coupling-mediated channel, which we explore in \secref{sec:dressing}.


\subsection{Effective time-independent Hamiltonian via a high-frequency expansion}
\label{sec:HFE}

Before we can analyze the coupling-mediated crosstalk, we need to transform the Hamiltonian of the driven system into a time-independent form. Hence, we apply a high-frequency expansion, while retaining corrections arising from the rapidly oscillating terms~\cite{Bukov2015, EckardtAnisimovas2015, goldman2014periodically}.

Without loss of generality, we consider driving only through the XY$_1$ control line ($V_{d2}=0$) at a single frequency $\Omega$, i.e., $V_{d1}(t)=V_{d1}\cos(\Omega t)$. Since each drive strength $\epsilon_k$ depends linearly on $V_{d1}$, all three drive strengths inherit the same time dependence, $\epsilon_k(t)=\epsilon_k\cos(\Omega t)$, with relative strengths $\lambda_{11}=1$, $\lambda_{21}=\epsilon_2/\epsilon_1$, and $\lambda_{c1}=\epsilon_c/\epsilon_1$. We only retain the leading Kerr nonlinearity $\tfrac{\alpha_k}{2} a_k^{\dag 2}a_k^2$, dropping the higher-order counter-rotating corrections in \eqref{eq:kerr_h}, which are subleading and do not affect the crosstalk analysis here. The resulting driven Hamiltonian is
\begin{align}
  H =&\ \sum_{k \in \{1,2,c\}}
  \mleft(
  \omega_k a_k^\dag a_k
  + \frac{\alpha_k}{2} a_k^{\dag 2} a_k^2
  \mright)
  \nonumber \\
  &+ \sum_{ij \in \{1c,2c,12\}} g_{ij} X_i X_j
  \nonumber \\
  &+ i\epsilon_1\cos(\Omega t)
  \mleft(
  \lambda_{11}X_1
  + \lambda_{21}X_2
  + \lambda_{c1}X_c
  \mright) \, ,
  \label{eq:DrivenH}
\end{align}
and the corresponding Hamiltonian for driving through XY$_2$ ($V_{d1}=0$) follows by interchanging the indices 1 and 2 in the drive terms.

To apply the high-frequency expansion, the relevant energy scales of the Hamiltonian must be small compared with the drive frequency $\Omega$~\cite{goldman2014periodically}. In the laboratory frame, this condition is not satisfied because the bare frequencies $\omega_k$ are comparable with $\Omega$. We therefore transform the Hamiltonian in \eqref{eq:DrivenH} to a frame rotating at $\Omega$, in which the bare frequencies are replaced by the detunings $\Delta_k=\omega_k-\Omega$. In the parameter regime considered here, these detunings $\Delta_k$, together with the coupling strengths $g_{ij}$ and anharmonicities $\alpha_k$, are small compared with $\Omega$. The applicability of the expansion additionally requires that the drive strengths satisfy $|\epsilon_k|\ll\Omega$. The transformed Hamiltonian is
\begin{widetext}
\begin{align}
  H' =&\ \sum_{k \in \{1,2,c\}} \mleft[ (\omega_k-\Omega) a_k^\dag a_k + \frac{\alpha_k}{2} a_k^\dag a_k^\dag a_k a_k \mright] 
  - \sum_{ij \in \{1c,2c,12\}} g_{ij} \mleft[ a_i^\dag a_j + a_i a_j^\dag \mright] 
  + \frac{i\epsilon_1}{2} \sum_{k \in \{1,2,c\}} \lambda_{k1} \mleft(  a_k^\dag -  a_k \mright) \nonumber \\
  &+ \sum_{ij \in \{1c,2c,12\}} g_{ij} \mleft[ e^{i2\Omega t} a_i^\dag a_j^\dag + e^{-i2\Omega t} a_i a_j \mright] 
  + \frac{i\epsilon_1}{2} \sum_{k \in \{1,2,c\}} \lambda_{k1} \mleft(  e^{i2\Omega t} a_k^\dag -  e^{-i2\Omega t} a_k \mright) \, .
  \label{eq:DrivenHRotating}
\end{align}
\end{widetext}
The first line is the static part of the transformed Hamiltonian, while the last line only contains terms rotating at twice the driving frequency.

We want to perturbatively eliminate the fast-rotating $2\Omega$ terms in \eqref{eq:DrivenHRotating} to obtain an effective, time-independent Hamiltonian retaining all three modes.
Since the high-frequency expansion relies on the separation between the drive frequency $\Omega$ and the other energy scales of the Hamiltonian, rather than on a large qubit-coupler detuning, it remains applicable to both far-detuned and near-resonant configurations. Applying the expansion to first order in $1/\Omega$, as detailed in \appref{app:magnus}, we obtain the effective time-independent Hamiltonian
\begin{align}
  H_{\rm eff}^{(1)} =&\ \sum_{k \in \{1,2,c\}} \mleft[ \mleft( \tilde\omega_k - \Omega \mright) a_k^\dag a_k + \frac{\alpha_k}{2} a_k^\dag a_k^\dag a_k a_k \mright] \nonumber \\
  &- \sum_{ij \in \{1c,2c,12\}} \tilde g_{ij} \mleft(a_i^\dag a_j + a_i a_j^\dag \mright) \nonumber \\
  &+ \sum_{k \in \{1,2,c\}} \mleft( \frac{i\epsilon_1}{2} \tilde\lambda_{k1}a_k^\dag + \text{H.c.} \mright) + \mathcal{O}(1/\Omega^2) \, .
  \label{eq:heff1_final}
\end{align}
Here, the tildes indicate that the frequencies, coupling strengths, and drive strengths have been renormalized:
\begin{align}
  \tilde\omega_k =&\ \omega_k - \frac{g_{ik}^2+g_{jk}^2}{2\Omega} \, ,
  \label{eq:tilde_omega}
\\
  \tilde g_{ij} =&\ g_{ij}+\frac{g_{ik}g_{jk}}{2\Omega} \, ,
  \label{eq:tilde_g}
\\
  \tilde\lambda_{k1} =&\ \lambda_{k1}+\frac{g_{ki}\lambda_{i1}+g_{kj}\lambda_{j1}}{2\Omega} \, , 
  \label{eq:tilde_Lambda}
\end{align}
where $i$, $j$, and $k$ are distinct in each expression. Since $\tilde\omega_k$ and $\tilde g_{ij}$ only depend on the static circuit and $\Omega$, they are unaffected by which XY control line is used for the drive; if we drive through XY$_2$ instead, only the drive term changes with $\epsilon_1 \to \epsilon_2$ and $\tilde\lambda_{k1} \to \tilde \lambda_{k2}$.


\subsection{Full crosstalk in the dressed basis}
\label{sec:dressing}

We now take the final steps from the time-independent effective Hamiltonian in \eqref{eq:heff1_final} to a full analytical picture of the crosstalk. To do so, we note that the linear part of $H_{\rm eff}^{(1)}$ in \eqref{eq:heff1_final} still couples the three modes through the $\tilde g_{ij}$ terms on the second line. As such, $a_1$, $a_2$, and $a_c$ are not the true normal modes of the static Hamiltonian: driving $a_1$ will also excite $a_2$ and $a_c$ through this residual coupling. There are thus two distinct crosstalk mechanisms: a coupling-mediated channel from $\tilde g_{ij}$ and a drive-mediated channel captured by $\tilde\lambda_{k1}$. 

Clarifying the effect of the coupling-mediated crosstalk requires finding the normal modes of the coupled system. Since the high-frequency expansion in \secref{sec:HFE} removed all pairing terms $a_i^\dag a_j^\dag$ and $a_ia_j$, diagonalizing the coupling-mediated channel only requires a unitary rotation among the annihilation operators, i.e., a Bogoliubov transformation. In \appref{app:diagonalization}, we perform this diagonalization to obtain the effective Hamiltonian in terms of the normal modes $b_1$, $b_2$, and $b_c$. The result is
\begin{align}
  H_{\rm eff, d}^{(1)} =&\ \sum_{m \in \{1,2,c\}} \mleft[ \mleft( \omega_{{\rm eff},m} - \Omega \mright) b_m^\dag b_m + \frac{\bar\alpha_m}{2} b_m^\dag b_m^\dag b_m b_m \mright] \nonumber \\
  &+ \sum_{mm' \in \{12,1c,2c\}} \chi_{mm'} b_m^\dag b_m b_{m'}^\dag b_{m'} \nonumber \\
  &+ \frac{i\epsilon_1}{2} \sum_{m \in \{1,2,c\}} \Theta_m\mleft( b_m^\dag -  b_m \mright) \, .
  \label{eq:heff_dressed_final}
\end{align}
The full expressions for the effective frequencies $\omega_{{\rm eff},m}$, the effective anharmonicities $\bar\alpha_m$, and the effective cross-Kerr couplings $\chi_{mm'}$ in \eqref{eq:heff_dressed_final} are given in \appref{app:diagonalization}. 

For modeling the crosstalk, the key quantity in \eqref{eq:heff_dressed_final} is the last line, where the dressed drive amplitude takes the form
\begin{align}
  \Theta_m =&\ \frac{1}{|v_m|}\Big[\tilde\lambda_{11}\mleft(\tilde g_{1c}\mleft(\tilde\omega_2-\omega_{{\rm eff},m}\mright)+\tilde g_{12}\tilde g_{2c}\mright) \nonumber \\
  &+ \tilde\lambda_{21}\mleft(\tilde g_{2c}\mleft(\tilde\omega_1-\omega_{{\rm eff},m}\mright)+\tilde g_{12}\tilde g_{1c}\mright) \nonumber \\
  &+ \tilde\lambda_{c1}\mleft(\mleft(\tilde\omega_1-\omega_{{\rm eff},m}\mright)\mleft(\tilde\omega_2-\omega_{{\rm eff},m}\mright)-\tilde g_{12}^2\mright)\Big] \, ,
  \label{eq:theta_explicit}
\end{align}
with $|v_m|$ denoting the norm of the eigenvector associated with the dressed mode $m$, as given in \appref{app:diagonalization}. Equation~(\ref{eq:theta_explicit}) shows that the dressed drive amplitude $\Theta_m$ is a coherent sum of three contributions proportional to the relative drive strengths $\tilde\lambda_{11}$, $\tilde\lambda_{21}$, and $\tilde\lambda_{c1}$. As the coupler is tuned, the coupler-related quantities $\tilde\lambda_{c1}$, $\tilde g_{1c}$, and $\tilde g_{2c}$ vary together with the dressed-mode frequencies $\omega_{{\rm eff},m}$, changing the relative weights of the three contributions. These contributions can therefore interfere constructively or destructively, with a crosstalk null occurring when they cancel for the victim mode, such that $\Theta_m=0$.


\subsection{Definitions of crosstalk quantities}
\label{sec:DefinitionsCrosstalk}

Using the dressed drive amplitudes in \eqref{eq:theta_explicit}, we now define a few quantities related to crosstalk, which we will use in the rest of the article. First, the Rabi frequency of qubit mode Q$_k$, driven through the XY$_i$ control line at frequency $\Omega$, is taken to be
\begin{equation}
  R_{i,k} (\Omega) = \epsilon_i \mleft| \Theta_k(\Omega) \mright| \, .
  \label{eq:omega_rabi}
\end{equation}
Note that this definition of the Rabi frequency is valid only in the single-excitation regime, as we have neglected Kerr effects.

The self-Rabi frequency of Q$_i$, when driven through its dedicated XY$_i$ control line at the resonance frequency $\omega_{\mathrm{eff},i}$, is defined as
\begin{equation}
  R^{\mathrm{self}}_i \equiv R_{i,i} (\omega_{\mathrm{eff},i})
  = \epsilon_i \mleft| \Theta_i(\omega_{\mathrm{eff},i}) \mright| \, .
  \label{eq:omega_self}
\end{equation}
For $i\neq j$, the cross-Rabi frequency of Q$_j$, when driven through the XY$_i$ control line at its resonance frequency $\omega_{\mathrm{eff},j}$, is defined as
\begin{equation}
  R^{\mathrm{cross}}_{i,j} \equiv R_{i,j}(\omega_{\mathrm{eff},j})
  = \epsilon_i \mleft| \Theta_j(\omega_{\mathrm{eff},j}) \mright| \, .
  \label{eq:omega_cross}
\end{equation}
Crosstalk between drive line XY$_i$ and qubit Q$_j$ is typically expressed as
\begin{equation}
    {\Lambda}_{i,j} =
    20 \log_{10} \mleft( \frac{R^{\rm cross}_{i,j}}{R^{\rm self}_i} \mright) \, .
    \label{eq:xtalk_dB}
\end{equation}

For clarity, we introduce a simplified notation for the self- and cross-Rabi frequencies in our two-qubit processor. The self-Rabi frequencies of Q$_1$ and Q$_2$, when driven through their respective drive lines XY$_1$ and XY$_2$, are denoted as
\[
R^{\rm self}_1 \rightarrow R^{\rm self}_{\rm Q1} \, ,
\qquad
R^{\rm self}_2 \rightarrow R^{\rm self}_{\rm Q2} \, .
\]
Similarly, the cross-Rabi frequencies of Q$_1$ and Q$_2$, when driven through the opposite drive lines XY$_2$ and XY$_1$, respectively, are denoted as
\[
R^{\rm cross}_{2,1} \rightarrow R^{\rm cross}_{\rm Q1} \, ,
\qquad
R^{\rm cross}_{1,2} \rightarrow R^{\rm cross}_{\rm Q2} \, .
\]

It is worth mentioning that the analytical model for the Rabi response is obtained using a high-frequency expansion truncated at first order. To assess the accuracy of this truncation, we compare the analytical predictions for the cross-Rabi frequencies with full numerical simulations in Appendix~\ref{app:analytics_vs_numerics}.


\section{Experiment}
\label{sec:Experiment}

We experimentally investigate crosstalk in the device shown in \figref{fig:device}, with the relevant parameters reported in \tabref{tab:floating_system_params} in \appref{app:exp_setup}. We measure both self-Rabi frequencies ($R^{\rm self}_{\rm Q1}$ and $R^{\rm self}_{\rm Q2}$) and cross-Rabi frequencies ($R^{\rm cross}_{\rm Q1}$ and $R^{\rm cross}_{\rm Q2}$) as a function of the coupler frequency. To establish a reference for the measurements, we first calibrate the drive power on each control line such that the corresponding target qubit has a self-Rabi frequency of \SI{20}{\mega\hertz} when the coupler is far detuned. We then keep the drive power fixed and measure the corresponding cross-Rabi frequencies, ensuring a consistent basis for quantifying the crosstalk. We probe on-resonance crosstalk, corresponding to the drive being resonant with the victim qubit and representing the worst-case scenario for gate operations. 

\begin{figure*}
    \centering
    \includegraphics[width=0.9\textwidth]{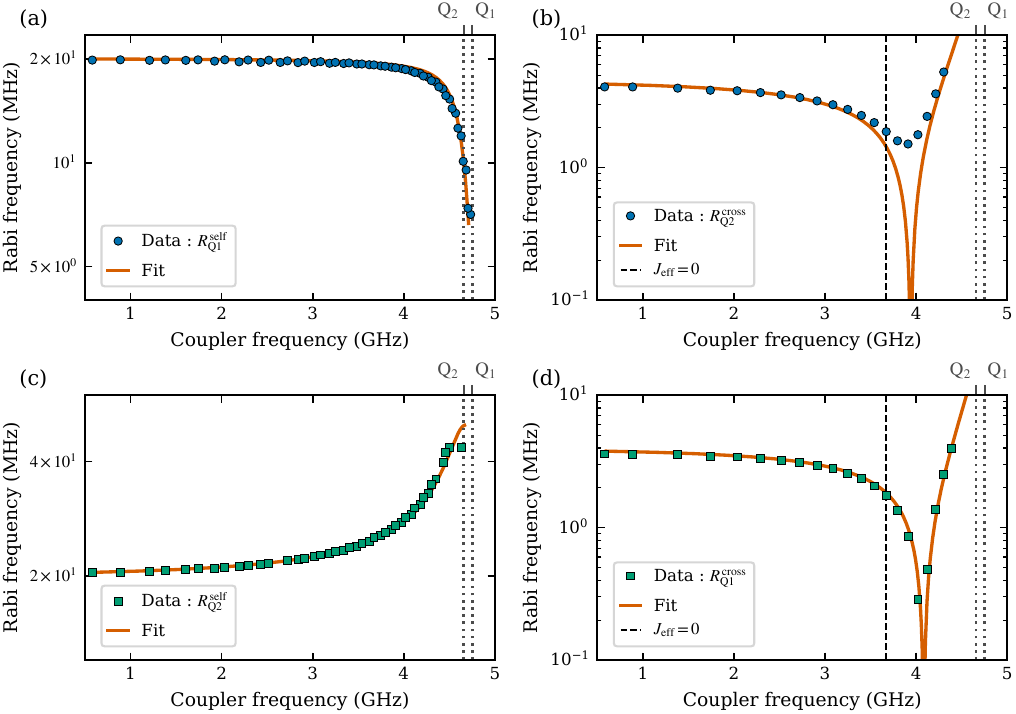}
     \vspace{-1.0ex}
\caption{Experimental self- and cross-Rabi frequencies as a function of coupler frequency.
(a, c) Measured self-Rabi rates of Q$_1$ driven through XY$_1$ and Q$_2$ driven through XY$_2$, respectively, as a function of coupler frequency. 
(b, d) Corresponding cross-Rabi rates of Q$_2$ driven through XY$_1$ and Q$_1$ driven through XY$_2$. At each coupler bias, the driven qubit is excited at its resonance frequency. Markers show the experimental data, while the solid curves show the fit given by Eqs.~(\ref{eq:omega_self}) and (\ref{eq:omega_cross}). The vertical dotted lines in all panels indicate the qubit frequencies. The vertical dashed line in (b, d) indicates the coupler frequency at which $J_{\mathrm{eff}} = 0$. 
    \label{fig:rabi_grid}}
\end{figure*}

In \figref{fig:rabi_grid}, data obtained by driving through XY$_1$ are shown as blue circles, while those obtained through XY$_2$ are represented by green squares. Figures~\figpanelNoPrefix{fig:rabi_grid}{a} and \figpanelNoPrefix{fig:rabi_grid}{c} show the measured self-Rabi frequencies $R^{\rm self}_{\rm Q1}$ and $R^{\rm self}_{\rm Q2}$, respectively. The corresponding cross-Rabi frequencies, $R^{\rm cross}_{\rm Q1}$ and $R^{\rm cross}_{\rm Q2}$, are shown in Figs.~\figpanelNoPrefix{fig:rabi_grid}{b} and \figpanelNoPrefix{fig:rabi_grid}{d}, respectively. In both cross-Rabi measurements, we observe a dip at a specific coupler frequency. This suppression indicates that the effective drive amplitude experienced by the victim qubit is strongly reduced at this operating point. 

One possible explanation for the suppression of the cross-Rabi frequencies is the cancellation of the effective qubit-qubit coupling at that given coupler frequency~\cite{zhao2022quantum, magesan2020effective}. We therefore compare the experimentally observed minima with the condition $J_{\mathrm{eff}} = 0$, where $J_{\mathrm{eff}}$ is the effective qubit-qubit coupling given by~\cite{yan2018tunable}
\begin{equation}
    J_\mathrm{eff} =
    g_{12} + g_{1c} g_{2c} \mleft( \frac{1}{\Delta_m} - \frac{1}{\Sigma_m} \mright) ,
\label{Jeff_eq}
\end{equation}
with
\begin{align}
    \frac{1}{\Delta_m}
    &=
    \frac{1}{2} \mleft( \frac{1}{\omega_1 - \omega_c} +
    \frac{1}{\omega_2 - \omega_c} \mright) \, ,
    \\
    \frac{1}{\Sigma_m}
    &=
    \frac{1}{2} \mleft( \frac{1}{\omega_1 + \omega_c} +
    \frac{1}{\omega_2 + \omega_c} \mright) \, .
\end{align}

For our device, we extract the values of the couplings $g_{ij}$ by fitting the Hamiltonian model to the measured qubit spectroscopies and their ZZ interaction, reported in Fig.~\ref{fig:qubit_spec} and Fig.~\ref{fig:ZZ} in \appref{app:spectro}  and \appref{app:ZZ}, respectively. We estimate $J_{\mathrm{eff}} = 0$ to occur at a coupler frequency of \SI{3.665}{\giga\hertz}, indicated by the dashed vertical line in Figs.~\figpanelNoPrefix{fig:rabi_grid}{b} and
\figpanelNoPrefix{fig:rabi_grid}{d}. This frequency is approximately \SI{0.3}{\giga\hertz} below the observed minima in the cross-Rabi frequencies. Thus, the observed suppression of the cross-Rabi frequency cannot be attributed to the cancellation of the effective qubit-qubit coupling alone. The full dependence of $J_{\mathrm{eff}}$ on the coupler frequency is shown in \figref{fig:J_eff} in \appref{app:Jeff}.

We therefore consider additional drive pathways present in the device. The microwave drive can couple not only to the target qubit, but also directly to the coupler and to the victim qubit, as shown in \figref{fig:circuit_schematic}. We model these contributions utilizing the capacitance model introduced in \secref{sec:model}. For each drive line, we simultaneously fit the corresponding self- ($R^{\rm self}_{\rm Q1(2)}$) and cross-Rabi ($R^{\rm cross}_{\rm Q2(1)}$) measurements using Eqs.~(\ref{eq:omega_self}) and (\ref{eq:omega_cross}). The design values of the drive-to-node capacitances listed in \tabref{tab:drive_model_hierarchy} are used as initial fitting parameters. For the cross-Rabi fitting, we additionally introduce an amplitude scaling factor, $k$, to account for variations in attenuation in the cryostat wiring, i.e., $R^{\mathrm{cross}}_{i,j} \equiv k R_{i,j}(\omega_{\mathrm{eff},j})$.

\begin{table}
\centering
\caption{Designed and extracted values for the drive-to-node capacitances, given in aF, and amplitude scaling factors $k$ used in the Rabi-frequency analysis.}
\label{tab:drive_model_hierarchy}

\renewcommand{\arraystretch}{1.1}
\resizebox{\columnwidth}{!}{%
\begin{tabular}{llccc c}
\hline\hline
\textbf{Drive} & \textbf{Capacitance} & \textbf{Parameter} & \textbf{Design} & \textbf{Fit} & \textbf{$k$} \\
\hline
\multirow{3}{*}{$\rm{XY}_1$} 
  & Target & $C_{d1u,1}, C_{d1d,1}$ & 80, 157 & 96.6, 145.4 & \multirow{3}{*}{1.911} \\
  & Coupler   & $C_{dc,1}$             & 19       & 18.4       &  \\
  & Victim    & $C_{d2u,1}, C_{d2d,1}$ & 6, 5    & 2.3, 7.7    &  \\
\hline
\multirow{3}{*}{$\rm{XY}_2$} 
  & Target & $C_{d2u,2}, C_{d2d,2}$ & 152, 68 & 122.1, 105.6 & \multirow{3}{*}{0.420} \\
  & Coupler   & $C_{dc,2}$             & 30      & 29.5        &  \\
  & Victim    & $C_{d1u,2}, C_{d1d,2}$ & 11, 7   & 2.5, 10.3   &  \\
\hline\hline
\end{tabular}%
}
\end{table}

The resulting fits are shown as solid lines in each plot in \figref{fig:rabi_grid}. Including both the parasitic drive-to-coupler and drive-to-victim capacitances yields excellent agreement with the measured self- and cross-Rabi frequencies, except for the data in Fig.~\figpanelNoPrefix{fig:rabi_grid}{b} where we do observe a smaller pronounced dip. The capacitance values, extracted from the fit, are summarized in \tabref{tab:drive_model_hierarchy}; deviations from the design values may be partially attributed to fabrication, although other effects, such as electromagnetic modes in the sample holder may also contribute. In contrast, excluding the additional drive pathways does not reproduce the observed frequency dependence; see \appref{app:comparison_models} for details.

\begin{figure}
    \centering
    \includegraphics[width=0.45\textwidth]{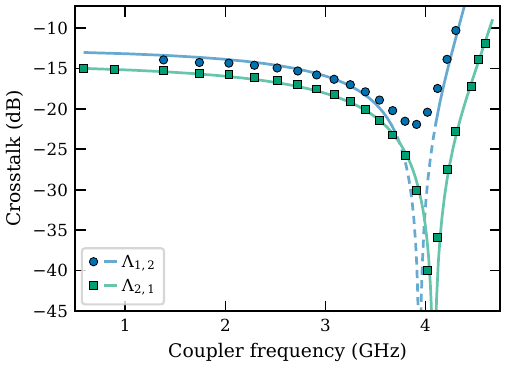}
\caption{Measured crosstalk for the two drive lines as a function of coupler frequency, expressed in dB using \eqref{eq:xtalk_dB}. The quantities $\Lambda_{2,1}$ and $\Lambda_{1,2}$ correspond to crosstalk from $\rm{XY}_1$ to $\rm{Q}_2$ and from $\rm{XY}_2$ to $\rm{Q}_1$, respectively. More negative values indicate stronger isolation between the drive line and the victim qubit.  The dashed blue line indicates the
stronger suppression of $\Lambda_{1,2}$ predicted
by the analytical model, which is not observed
experimentally. This discrepancy may arise from
additional crosstalk mechanisms not accounted for
in the model.}
    \label{fig:selectivity_plot}
\end{figure}

The resulting crosstalk ratios, $\Lambda_{1,2}$ and $\Lambda_{2,1}$, defined in \eqref{eq:xtalk_dB}, are shown in \figref{fig:selectivity_plot} as a function of coupler frequency, with blue circles and green squares corresponding to $\Lambda_{1,2}$ and $\Lambda_{2,1}$, respectively. The measured values are in good agreement with the model (solid lines) resulting from the fitting in \figref{fig:rabi_grid}. This agreement demonstrates that the frequency dependence of the crosstalk, including its pronounced suppression, can be quantitatively captured by the additional drive pathways included in our model. In particular, the model identifies an operating point at which the contributions from the different drive pathways strongly suppress the effective drive amplitude at the victim qubit, resulting in a minimum in the crosstalk. The observed discrepancy in  $\Lambda_{1,2}$ likely arises from additional microwave-crosstalk contributions not included in the present circuit model.

For completeness, we also report the full self- and cross-Rabi dynamics as a function of coupler magnetic flux in Appendix~\ref{app:exp_vs_simulation_2D}, showing good agreement between experimental data and simulations.


\section{Discussion}
\label{sec:Discussion}

The model in  \secref{sec:model}, which incorporates parasitic drive capacitances, accurately predicts the experimental crosstalk observed in \secref{sec:Experiment}. We now turn to investigate design considerations and guidelines that follow from the success of this model. First, we compare the role of the parasitic drive capacitances in architectures with floating and grounded qubits. Then, we consider off-resonant crosstalk, which is more common than the worst-case resonant crosstalk measured in \secref{sec:Experiment}, and show how the two are related.


\subsection{Role of parasitic drive capacitances}

A key finding of this work is that parasitic drive capacitances play an important role in the suppression of drive crosstalk. We demonstrated above that interference between the different drive pathways gives rise to a coupler frequency at which the cross-Rabi rate is minimized. We refer to this operating point as the crosstalk-null frequency.

We now compare the role of the parasitic drive capacitances in floating- and grounded-qubit architectures. The model in \secref{sec:model} describes a floating-qubit architecture. We derive the corresponding circuit model and relevant parameters for the grounded case in \appref{app:grounded_qubit}. 

For floating qubits, we define the differential drive-to-victim capacitances $\Delta C_{\mathrm{victim},i}$ as the difference of the capacitance between the drive line and the two pads of the victim qubit: 
$\Delta C_{\mathrm{victim},1} = C_{d2u,1} - C_{d2d,1}$ and 
$\Delta C_{\mathrm{victim},2} = C_{d1u,2} - C_{d1d,2}$, 
with the capacitance values reported in \tabref{tab:drive_model_hierarchy}. In the case of grounded qubits, we define 
$C_{\mathrm{victim}}=C_{d1,2}=C_{d2,1}$ 
as the capacitance from one drive line to its victim qubit. We use similar capacitance values when performing simulations: a drive-to-target capacitance of \SI{100}{\atto\farad}, a drive-to-coupler capacitance of \SI{20}{\atto\farad}, and a drive-to-victim capacitance of \SI{5}{\atto\farad}.

\begin{figure*}
    \centering
    \includegraphics[width=\textwidth]
    {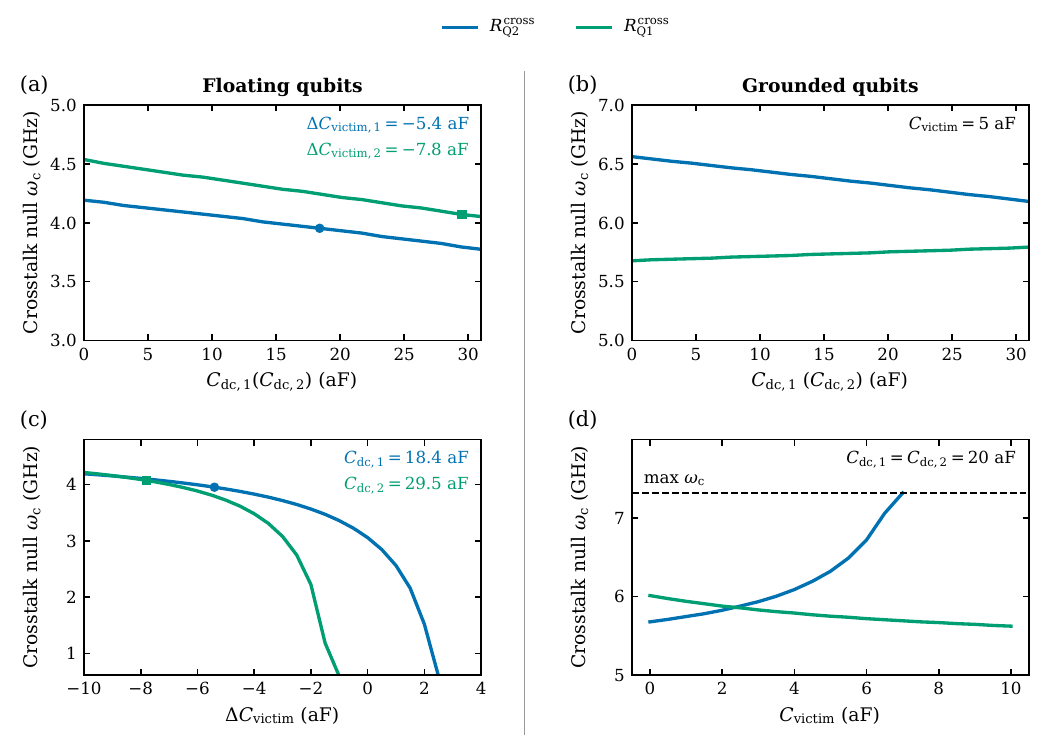}
    \vspace{-4.4ex}
\caption{Dependence of the crosstalk-null frequencies of $R^{\rm cross}_{\rm Q2}$ (blue) and $R^{\rm cross}_{\rm Q1}$ (green) on the parasitic drive capacitances.
(a, b) Crosstalk-null frequencies for Q$_2$ driven through XY$_1$ and Q$_1$ driven through XY$_2$ as a function of the drive-to-coupler capacitance for the floating- and grounded-qubit architectures, respectively.
(c, d) Corresponding crosstalk-null frequencies as a function of the drive-to-victim capacitance for the floating- and grounded-qubit architectures, respectively. The blue circles and green squares in (a, c) indicate the fitted capacitance values from \tabref{tab:drive_model_hierarchy} for XY$_1$ and XY$_2$, respectively. The horizontal dashed line in (d) marks the maximum coupler frequency.}
     \label{fig:xtalk_nulls}
\end{figure*}

Figures~\figpanelNoPrefix{fig:xtalk_nulls}{a} and \figpanelNoPrefix{fig:xtalk_nulls}{b} show the crosstalk-null frequency for the floating- and grounded-qubit case, respectively. The crosstalk-null frequency is shown as a function of the drive-to-coupler capacitances $C_{dc,1}$ and $C_{dc,2}$, for the cross-Rabi frequencies $R^{\rm cross}_{\rm Q2}$ (blue) and $R^{\rm cross}_{\rm Q1}$ (green). For floating qubits, the two crosstalk-null frequencies remain relatively close and vary with slopes of the same sign as the drive-to-coupler capacitance is changed. This behavior provides flexibility to bring the two frequencies to coincide to achieve complete cancellation of control crosstalk through suitable choices of drive-to-coupler capacitances. In other words, we can draw a horizontal line in \figpanel{fig:xtalk_nulls}{a} such that it goes through both the blue and the green curves, corresponding to a single crosstalk-null frequency set by different $C_{dc,1}$ and $C_{dc,2}$. In contrast, for grounded qubits, the two null frequencies change with slopes of opposite sign and are more widely separated over the considered range. Consequently, a larger adjustment of the drive-to-coupler capacitance is required to eliminate control crosstalk.   

In Figs.~\figpanelNoPrefix{fig:xtalk_nulls}{c} and \figpanelNoPrefix{fig:xtalk_nulls}{d}, we show, for a given value of the drive-to-coupler capacitance, the crosstalk-null frequencies as a function of the drive-to-victim capacitances, for the floating and grounded-qubit case, respectively. Here, for floating qubits, we have kept $\Delta C_{\mathrm{victim}} = \Delta C_{\mathrm{victim},1} = \Delta C_{\mathrm{victim},2}$. In \figpanel{fig:xtalk_nulls}{c}, we see that as $\Delta C_{\mathrm{victim}}$ increases in negative values, the dependence of the crosstalk-null frequency on $\Delta C_{\mathrm{victim}}$ becomes weaker, thus reducing the precision required in fabrication to attain the desired capacitance and crosstalk cancellation with floating qubits. On the other hand, for grounded qubits, the two crosstalk-null frequencies shift in opposite directions with $C_{\mathrm{victim}}$ in \figpanel{fig:xtalk_nulls}{d}; they coincide only at a particular capacitance value, making simultaneous alignment more sensitive to fabrication variations.

Finally, we remark that the crosstalk-null trends discussed for floating qubits depend on the chosen drive-to-target island capacitance asymmetry, $C_{d1u,1} - C_{d1d,1}$ for XY$_1$ and $C_{d2u,2} - C_{d2d,2}$ for XY$_2$, and therefore do not represent a general property of all floating-qubit designs. In Appendix~ \ref{app:victim_sum_cap} we report, for completeness, the dependence of the crosstalk-null frequency on the common-mode drive-to-victim capacitance $\Sigma C_{\mathrm{victim}}$.

Overall, these results show that the drive-to-coupler and drive-to-victim capacitances provide useful parameters for controlling the crosstalk-null frequency, particularly for floating-qubit architectures. Although often neglected in standard circuit designs, these parasitic capacitances can serve as practical control knobs for suppressing microwave drive crosstalk.


\subsection{Off-resonance crosstalk dynamics}

The resonant crosstalk considered so far only provides a measure of the worst-case crosstalk. 
During quantum gate operations, however, the control signal is resonant with the target qubit, and the residual microwave crosstalk drives the victim qubit off-resonantly. This off-resonant drive causes unwanted excitation and constitutes one of the primary error mechanisms in simultaneous single-qubit gate operations, limiting overall gate fidelity. We therefore estimate the resulting maximum excited-state population of the victim qubit as a function of the coupler frequency in a realistic setting.

We consider a drive applied through XY$_i$ at the resonance frequency of the target qubit Q$_i$, such that $\Omega = \omega_{\mathrm{eff},i}$. For $i \neq j$, the victim qubit Q$_j$ is detuned from the drive by
\begin{equation}
  \Delta_{i,j} \equiv
  \omega_{\mathrm{eff},j} -
  \omega_{\mathrm{eff},i} \, .
  \label{eq:delta_victim}
\end{equation}
The corresponding off-resonant Rabi frequency of Q$_j$ is
\begin{equation}
  R_{i,j}(\omega_{\mathrm{eff},i}) =
  \epsilon_i \mleft| \Theta_j(\omega_{\mathrm{eff},i}) \mright| \, .
\end{equation}

In the weak-drive limit,
$R_{i,j}(\omega_{\mathrm{eff},i}) \ll |\bar{\alpha}_j|, |\Delta_{i,j} - \bar{\alpha}_j|$,
we restrict Q$_j$ to the $\{\ket{0}, \ket{1}\}$ subspace. Neglecting small cross-Kerr shifts and choosing the $x$ axis along the effective drive direction in the transverse plane, the Hamiltonian of the victim qubit reduces to
\begin{equation}
  H_j^{\rm two\text{-}level} =
  \frac{\Delta_{i,j}}{2} \sigma_z +
  \frac{R_{i,j}(\omega_{\mathrm{eff},i})}{2} \sigma_x \, .
  \label{eq:H_victim_2level}
\end{equation}

For a victim qubit Q$_j$ initialized in the ground state $\ket{0}$, the Hamiltonian in \eqref{eq:H_victim_2level} gives the excited-state population
\begin{equation}
  P_j(t) =
  \frac{R_{i,j}^2(\omega_{\mathrm{eff},i})}{\widetilde{R}_{i,j}^{\,2}}
  \sin^2 \mleft( \frac{\widetilde{R}_{i,j} t}{2} \mright) \, ,
  \label{eq:generalized_rabi}
\end{equation}
where
\begin{equation}
  \widetilde{R}_{i,j} \equiv
  \sqrt{ R_{i,j}^2(\omega_{\mathrm{eff},i}) + \Delta_{i,j}^2}
\end{equation}
is the generalized Rabi frequency. The maximum crosstalk-induced excited-state population of Q$_j$ is therefore
\begin{equation}
  P_j^{\rm max} =
  \frac{R_{i,j}^2(\omega_{\mathrm{eff},i})}{\widetilde{R}_{i,j}^{\,2}} \, .
  \label{eq:pmax}
\end{equation}
%

\begin{figure}
  \centering
  \includegraphics[width=\columnwidth]{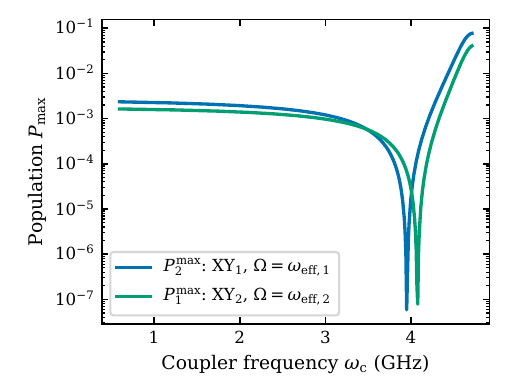}
\caption{Off-resonant crosstalk-induced population of the victim qubit as a function of coupler frequency, quantified by the maximum excited-state population $P_2^{\rm max}$ for Q$_1$ driven resonantly through XY$_1$ and $P_1^{\rm max}$ for Q$_2$ driven resonantly through XY$_2$. The minima correspond to the respective crosstalk-null frequencies.}
  \label{fig:crosstalk_null}
\end{figure}

The maximum population in \eqref{eq:pmax} provides an estimate of the unwanted excitation of the victim qubit during off-resonant driving. Figure~\ref{fig:crosstalk_null} shows the maximum population in $\mathrm{Q}_2$ ($P_2^{\rm max}$, blue) when Q$_1$ is driven resonantly through XY$_1$ and the maximum population in Q$_1$ ($P_1^{\rm max}$, green) when Q$_2$ is driven resonantly through XY$_2$. In both cases, the population exhibits a pronounced minimum near the corresponding crosstalk-null frequency, where the unwanted excitation is suppressed by several orders of magnitude. Away from the null, $P_j^{\rm max}$ increases rapidly to substantially higher values. Thus, the crosstalk null identified from the resonant response also corresponds to a strong suppression of the off-resonant population error, which is relevant during simultaneous single-qubit operations.


\section{Conclusion}
\label{sec:Conclusion}

We investigated classical microwave drive crosstalk in a superconducting quantum processor consisting of two floating transmon qubits coupled via a tunable coupler. Using a circuit model that accounts for parasitic capacitive coupling of the drive lines to both the coupler and the victim qubit, we derived analytical expressions for the crosstalk as a function of coupler frequency. Experimentally, we observed clear minima in the crosstalk-induced Rabi rates, consistent with our theoretical predictions. 

We showed that the suppression of drive crosstalk does not arise from cancellation of the effective qubit-qubit coupling alone, but also from destructive interference between multiple drive pathways enabled by the parasitic drive capacitances. Remarkably, this interference can be controlled by engineering the parasitic capacitances, which are typically neglected in circuit design. Achieving this control appears easier for floating qubits than for grounded ones.

In addition to classical drive crosstalk, the qubits are subject to static $ZZ$ interactions~\cite{fors2024comprehensive, chu2021coupler}, commonly referred to as quantum crosstalk, which can be suppressed by tuning the coupler frequency. Based on our insights, the parasitic drive capacitances could, in principle, be engineered to suppress classical microwave crosstalk at a low-$ZZ$ operating point. This suggests a device-level design strategy for simultaneously suppressing two distinct sources of coherent error at a common coupler operating frequency.

Overall, our findings point toward a design-based approach to mitigating microwave crosstalk, which could complement or replace more cumbersome existing compensation techniques.


\begin{acknowledgments}

We thank Nathan Goldman and Hang-Xi Li for useful discussions. Device fabrication was performed at Myfab Chalmers. Computational resources were provided by the National Academic Infrastructure for Supercomputing in Sweden (NAISS), funded by the Swedish Research council. Claude Sonnet (Anthropic) was used to assist with the development and refinement of scripts for fitting routines.
This research was financially supported by the Knut and Alice Wallenberg Foundation through the Wallenberg Center for Quantum Technology (WACQT) and the Horizon Europe programme HORIZON-CL4-2022-QUANTUM-01-SGA via the project 101113946 OpenSuperQPlus100. A.F.K.~is also supported by the Swedish Foundation for Strategic Research (grant numbers FFL21-0279 and FUS21-0063) and the Norwegian Research Council through the Norwegian Quantum Software Center (NorQSoft, project number 361350).

\end{acknowledgments}


\section*{Data Availability}

See the Appendices for theoretical and experimental details. All relevant data and figures supporting the main conclusions of the document are available on Zenodo \cite{aggarwal_2026_microwave_crosstalk}. Please refer to Anuj Aggarwal at anuja@chalmers.se if needed.


\section*{Conflict of Interest}

The authors have no conflicts of interest to disclose.


\section*{Author contributions}

A.A.~conceptualized the problem, designed the chip, performed the measurements, and analyzed the results, while S.P.F.~determined the device Hamiltonian parameters. A.A.~and J.F.P.~developed the theory, to which S.P.F.~contributed. A.A., M.R., A.O., J.B., A.F.R.~and A.N.~participated in the fabrication of the device. Z.M.~and H.J.~assisted with the measurements. J.F.P.~provided additional guidance during the project. A.A., A.F.K., and G.T.~wrote the manuscript. G.T.~led the supervision of the project, with contributions from A.F.K.~on the theory. All authors contributed to discussions and the interpretation of the results.


\appendix

\section{Experimental setup}
\label{app:exp_setup}

\begin{figure}
    \centering
    \includegraphics[width=\columnwidth]{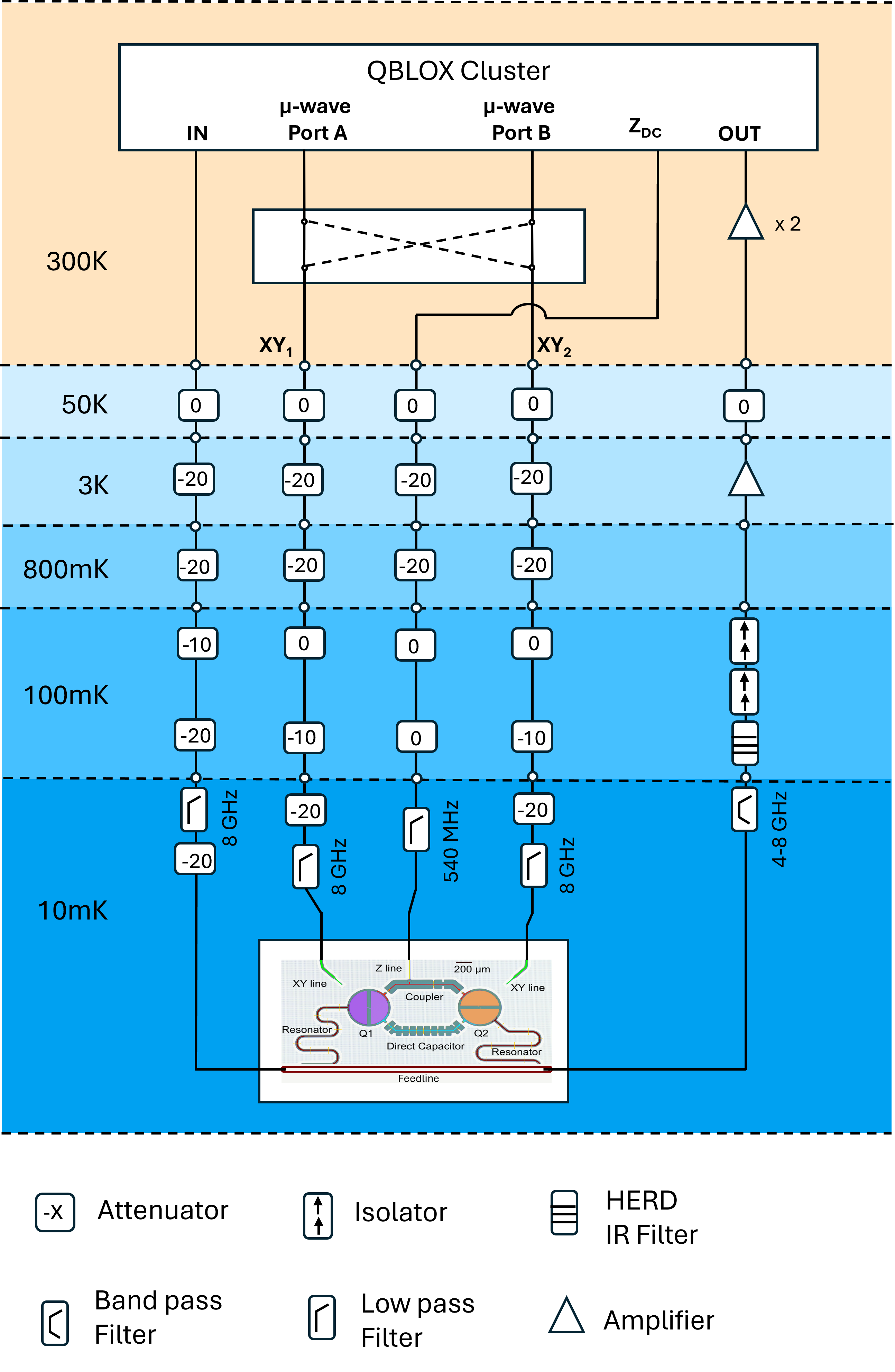}
    \caption{Detailed wiring schematic of the experimental setup. The device image at the bottom of the schematic is shown in more detail in \figref{fig:device}.
    \label{fig:wiring diagram}}
\end{figure}

An overview of the experimental setup and cryogenic wiring is shown in \figref{fig:wiring diagram}. The device is mounted on the \SI{10}{\milli\kelvin} mixing-chamber stage of a Bluefors LD250 dilution refrigerator. The microwave input lines are attenuated at successive temperature stages to suppress thermal photon noise reaching the device. The readout return line is filtered using HERD infrared filters~\cite{rehammar2023low} and isolated using directional isolators. The readout signal is first amplified by a high-electron-mobility transistor (HEMT) amplifier at the \SI{4}{\kelvin} stage, followed by two additional amplification stages at room temperature.

For each qubit, the same microwave source settings are used for the self- and cross-Rabi measurements, with only the drive line changed. For measurements of Q$_1$ using microwave port A, the signal is applied through XY$_1$ for the self-Rabi measurement and through XY$_2$ for the cross-Rabi measurement. Similarly, for measurements of Q$_2$ using microwave port B, the signal is applied through XY$_2$ for the self-Rabi measurement and through XY$_1$ for the cross-Rabi measurement.

Room-temperature control and data acquisition are provided by a Qblox Cluster equipped with QCM-RF, QRM-RF, and QCM modules. Single-qubit control pulses are generated by the QCM-RF modules and applied to the device through the XY drive lines. Readout pulses are sent to the device along the IN line, while the returning signal is digitized by the QRM-RF module from the OUT line. The DC flux bias for the tunable coupler is supplied by the QCM module through the $Z_{\mathrm{DC}}$ line.

The floating-transmon device, shown in \figref{fig:device}, is fabricated on a high-resistivity silicon substrate using aluminum thin-film processing, with the Josephson junctions defined by double-angle evaporation. The fabrication process is described in Refs.~\cite{aggarwal2025mitigating, biznarova2024mitigation, chayanun2024characterization}. The experimentally extracted Hamiltonian parameters are summarized in \tabref{tab:floating_system_params}; we extract the values of the couplings $g_{ij}$ by fitting the Hamiltonian model to the measured qubit spectroscopies and their ZZ interaction, reported in Fig.~\ref{fig:qubit_spec} and Fig.~\ref{fig:ZZ} in \appref{app:spectro}  and \appref{app:ZZ}, respectively. The designed and extracted capacitance values (see \figref{fig:circuit_schematic} for definitions) are given in \tabref{tab:designed_vs_fitted_caps}.

\begin{table}
\centering
\caption{System parameters for the floating-transmon device shown in \figref{fig:device}, including the two qubits Q$_1$ and Q$_2$, the tunable coupler C ($\Phi_{\rm{ext}}=0$), and their mutual coupling parameters.}
\label{tab:floating_system_params}

\renewcommand{\arraystretch}{1.1}
\resizebox{\columnwidth}{!}{%
\begin{tabular}{lccc}
\hline\hline
\textbf{Qubit Parameters} & \textbf{Q$_1$} & \textbf{Q$_2$} &  C \\
\hline
Frequency, $\omega_k / 2\pi$ (GHz)         & $4.752$  & $4.658$ & $4.93$ \\
Anharmonicity, $\alpha_k / 2\pi$ (GHz)     & $-0.24$ & $-0.24$  & $-0.144$ \\
Josephson energy, $E_{Jk} / h$ (GHz)       & $14.473$ & $13.944$  & $23.9$ \\
Charging energy, $E_{C,kk} / h$ (GHz)      & $0.214$& $0.214$& $0.135$ \\
\hline
\textbf{Coupling Parameters} & \textbf{Q$_1$-Q$_2$} & \textbf{Q$_1$-C} & \textbf{Q$_2$-C} \\
\hline
Coupling strength, $g_{ij} / 2\pi$ (GHz)   & $-0.0039$    & $0.077$  & $0.080$ \\
Cross-charging, $E_{C,ij} / h$ (GHz)       & $-0.00033$ & $0.0053$ & $0.0055$ \\
\hline\hline
\end{tabular}%
}
\end{table}

\begin{table}
\centering
\caption{Designed and extracted capacitance values.}
\label{tab:designed_vs_fitted_caps}

\renewcommand{\arraystretch}{1.1}
\renewcommand{\tabcolsep}{0.2cm}
\begin{tabular}{lcc}
\hline\hline
\textbf{Parameter} & \textbf{Designed (fF)} & \textbf{Extracted (fF)} \\
\hline
$C_{q1}$   & 29.0  & 36.9 \\
$C_{q2}$   & 28.8  & 36.5 \\
$C_{qc}$   & 110.0 & 129.0 \\
$C_{1c}$   & 7.66  & 7.70 \\
$C_{2c}$   & 7.80  & 7.52 \\
$C_{12}$   & 1.24  & 0.99 \\
$C_{1u,g}$ & 89.5  & 108.0 \\
$C_{1d,g}$ & 98.4  & 100.0 \\
$C_{2u,g}$ & 96.5  & 102.0 \\
$C_{2d,g}$ & 91.3  & 106.0 \\
\hline\hline
\end{tabular}
\end{table}


\section{From Lagrangian to Hamiltonian}
\label{app:LagrangianToHamiltonian}

Here, we give the details for the derivation in \secref{sec:CircuitLagrangian} of the Hamiltonian in \eqref{eq:kerr_h}, starting from the Lagrangian in \eqref{eq:Lagrangian} based on the circuit schematic in \figref{fig:circuit_schematic}. We reproduce \eqref{eq:Lagrangian} here for convenience:
\begin{align}
  \mathcal{L}
  =&\ \frac{1}{2} \dot{\mathbf{\phi}}^T C_m^{(0)} \dot{\mathbf{\phi}} - \dot{\mathbf{\phi}}^T q_d^{(0)} \nonumber\\
  &+ E_{J1} \cos \mleft( \frac{\phi_{1u} - \phi_{1d}}{\varphi_0} \mright) \nonumber \\
  &+ E_{J2} \cos \mleft( \frac{\phi_{2u} - \phi_{2d}}{\varphi_0} \mright) \nonumber \\
  &+ E_{Jc} (\Phi_{\text{ext}}) \cos \mleft( \frac{\phi_c}{\varphi_0} \mright) ,
  \label{eq:Lagrangian_app}
\end{align}
with the capacitance matrix
\begin{equation}
C_m^{(0)} =
\begin{pmatrix}
C_{1\Sigma,u} & -C_{q1} & 0 & -C_{12} & -C_{1c} \\
-C_{q1} & C_{1\Sigma,d} & 0 & 0 & 0 \\
0 & 0 & C_{2\Sigma,u} & -C_{q2} & -C_{2c} \\
-C_{12} & 0 & -C_{q2} & C_{2\Sigma,d} & 0 \\
-C_{1c} & 0 & -C_{2c} & 0 & C_{c\Sigma}
\end{pmatrix} \, ,
\end{equation}
where
\begin{align}
C_{1\Sigma,u} &= C_{q1} + C_{1u,g} + C_{1c} + C_{12} \nonumber \\
              &\mathrel{\phantom{=}} +\: C_{d1u,1} + C_{d1u,2} \, , \\
C_{1\Sigma,d} &= C_{q1} + C_{1d,g} + C_{d1d,1} + C_{d1d,2} \, , \\
C_{2\Sigma,u} &= C_{q2} + C_{2u,g} + C_{2c} + C_{d2u,1} + C_{d2u,2} \, , \\
C_{2\Sigma,d} &= C_{q2} + C_{2d,g} + C_{12} + C_{d2d,1} + C_{d2d,2} \, , \\
C_{c\Sigma}   &= C_{qc} + C_{1c} + C_{2c} + C_{dc,1} + C_{dc,2} \, ,
\end{align}
and the charge-drive vector
\begin{equation}
  q_d^{(0)} = V_{d1}
  \begin{pmatrix}
    C_{d1u,1} \\
    C_{d1d,1} \\
    C_{d2u,1} \\
    C_{d2d,1} \\
    C_{dc,1}
  \end{pmatrix} + V_{d2}
  \begin{pmatrix}
    C_{d1u,2} \\
    C_{d1d,2} \\
    C_{d2u,2} \\
    C_{d2d,2} \\
    C_{dc,2}
  \end{pmatrix} \, .
\end{equation}
%


\subsection{Common and differential modes}

We begin by noting that only differential fluxes enter in the potential terms for the qubits in \eqref{eq:Lagrangian_app}. Motivated by this observation, we transform the Lagrangian to a new flux basis
\begin{equation}
  \phi_{\rm new} = S \phi
\end{equation}
using the linear basis-change matrix
\begin{equation}
S =
\begin{pmatrix}
1 &  1 &  0 &  0 & 0 \\
0 &  0 &  1 &  1 & 0 \\
1 & -1 &  0 &  0 & 0 \\
0 &  0 &  1 & -1 & 0 \\
0 &  0 &  0 &  0 & 1
\end{pmatrix} \, .
\label{eq:Smatrix}
\end{equation}
Explicitly, this transformation defines
\begin{align}
  \phi_{1\pm} =&\ \phi_{1u} \pm \phi_{1d} \, , \nonumber \\
  \phi_{2\pm} =&\ \phi_{2u} \pm \phi_{2d} \, ,
\end{align}
while the coupler node $\phi_c$ is left unchanged. We group the new fluxes as 
\begin{equation}
\phi_{\rm new} = \begin{pmatrix} \phi_+ \\ \phi_- \end{pmatrix} \, ,
\end{equation}
such that the common modes are collected in $\phi_+ = (\phi_{1+},\phi_{2+})^T$ and $\phi_c$ is included in the differential-mode sector $\phi_- = (\phi_{1-},\phi_{2-},\phi_c)^T$.

Since $\phi = S^{-1} \phi_{\rm new}$, invariance of the capacitive terms and the drive terms in the Lagrangian determines the capacitance matrix and the charge-drive vector in the new basis:
\begin{align}
  C_m &= \mleft(S^{-1}\mright)^{T} C_m^{(0)} S^{-1} \, , \label{eq:Cm_new} \\
  q_d &= \mleft(S^{-1}\mright)^{T} q_d^{(0)} \, . \label{eq:qd_new}
\end{align}
Partitioning $C_m$ and $q_d$ according to the common- and differential-mode block structure, the Lagrangian becomes
\begin{align}
\mathcal{L} =&\ \frac{1}{2}
\begin{pmatrix} \dot{\phi}_+ & \dot{\phi}_- \end{pmatrix}
\begin{pmatrix}
C_{++} & C_{+-} \\
C_{+-}^T & C_{--}
\end{pmatrix}
\begin{pmatrix} \dot{\phi}_+ \\ \dot{\phi}_- \end{pmatrix}
\nonumber \\
&- \begin{pmatrix} \dot{\phi}_+ & \dot{\phi}_- \end{pmatrix}
\begin{pmatrix} q_{d,+} \\ q_{d,-} \end{pmatrix}
\nonumber \\
&+ \sum_{k\in\{1,2,c\}} E_{Jk}
\cos \mleft( \frac{(\phi_-)_k}{\varphi_0} \mright) \, ,
\label{eq:LagrangianCommonDifferential}
\end{align}
where $C_{++}$, $C_{--}$, and $C_{+-}$ are, respectively, the common-mode, differential-mode, and cross-mode blocks of $C_m$ coupling the $\phi_+$ and $\phi_-$ sectors, while $q_{d,+}$ and $q_{d,-}$ are the common-mode and differential-mode blocks, respectively, of $q_d$.

Since $q_d^{(0)}$ is a sum of two independent drive-line contributions, and the transformation $q_d = (S^{-1})^T q_d^{(0)}$ is linear, each line contributes to $q_{d,\pm}$ independently. For each drive line $\ell\in\{1,2\}$, we define the common- and differential-mode combinations of its capacitances to Q$_1$ and Q$_2$, respectively,
\begin{align}
C_{d1\pm,\ell} &\equiv \frac{1}{2}\mleft(C_{d1u,\ell} \pm C_{d1d,\ell}\mright) , \\
C_{d2\pm,\ell} &\equiv \frac{1}{2}\mleft(C_{d2u,\ell} \pm C_{d2d,\ell}\mright) ,
\end{align}
in terms of which the charge-drive-vector sectors evaluate to
\begin{align}
  q_{d,+} =&\ V_{d1}\,(C_{d1+,1},\, C_{d2+,1})^T \nonumber\\  &+ V_{d2}\,(C_{d1+,2},\, C_{d2+,2})^T \, , \\
  q_{d,-} =&\ V_{d1}\,(C_{d1-,1},\, C_{d2-,1},\, C_{dc,1})^T \nonumber\\ 
  &+ V_{d2}\,(C_{d1-,2},\, C_{d2-,2},\, C_{dc,2})^T \, .
\end{align}

Continuing to analyze the rewritten Lagrangian in \eqref{eq:LagrangianCommonDifferential}, we note that since $\phi_+$ is cyclic, its conjugate charge $q_+$ is conserved, and $\dot\phi_+$ can be eliminated. The conjugate charges, $q \equiv \partial\mathcal{L} / \partial\dot\phi$, are
\begin{align}
  \begin{pmatrix} q_+ \\ q_- \end{pmatrix}
  = \begin{pmatrix} C_{++} & C_{+-} \\ C_{+-}^T & C_{--} \end{pmatrix}
  \begin{pmatrix} \dot{\phi}_+ \\ \dot{\phi}_- \end{pmatrix} - \begin{pmatrix} q_{d,+} \\ q_{d,-} \end{pmatrix} \, .
\end{align}
Solving the first row for $\dot\phi_+$ and substituting into the second eliminates the common-mode sector, giving
\begin{equation}
  \dot{\phi}_- = C_{\rm eff}^{-1}\mleft(q_- + q_{d,\rm eff}\mright) \,  ,
\end{equation}
where the effective capacitance matrix is
\begin{equation}
  C_{\rm eff} = C_{--} - C_{+-}^T C_{++}^{-1} C_{+-}
\end{equation}
and the effective charge drive is
\begin{equation}
  q_{d,\rm eff} = q_{d,-} - C_{+-}^T C_{++}^{-1} \mleft( q_+ + q_{d,+} \mright) \, .
\end{equation}

As $q_+$ is fixed only by the initial charging of the common mode and would be a static offset, we take $q_+ = 0$, giving
\begin{equation}
  q_{d,\rm eff}\Big|_{q_+=0} = V_{d1}
     \begin{pmatrix}
       \tilde{C}_{d1} \\
       \tilde{C}_{d2} \\
       \tilde{C}_{dc}
     \end{pmatrix} + V_{d2}
     \begin{pmatrix}
       \breve{C}_{d1} \\
       \breve{C}_{d2} \\
       \breve{C}_{dc}
     \end{pmatrix} \, ,
\end{equation}
with each entry splitting into a direct contribution, from the corresponding parasitic capacitance, and a contribution induced by eliminating the common-mode sector. Since the two drives add linearly, the XY$_1$ line's contribution follows by setting $V_{d2} = 0$:
\begin{subequations}
\label{eq:effective_drive_caps}
\begin{align}
  \tilde{C}_{d1} V_{d1}
  =&\ \underbrace{C_{d1-,1} V_{d1}}_{\text{direct}}
  - \underbrace{\mleft[ C_{+-}^T C_{++}^{-1} q_{d,+} \mright]_{1-}}_{\text{induced}} \, ,
  \label{eq:effective_drive_cap_d1}
  \\
  \tilde{C}_{d2} V_{d1}
  =&\ \underbrace{C_{d2-,1} V_{d1}}_{\text{direct}}
  - \underbrace{\mleft[ C_{+-}^T C_{++}^{-1} q_{d,+} \mright]_{2-}}_{\text{induced}} \, ,
  \label{eq:effective_drive_cap_d2}
  \\
  \tilde{C}_{dc} V_{d1}
  =&\ \underbrace{C_{dc,1} V_{d1}}_{\text{direct}}
  - \underbrace{\mleft[ C_{+-}^T C_{++}^{-1} q_{d,+} \mright]_{c}}_{\text{induced}} \, .
  \label{eq:effective_drive_cap_dc}
\end{align}
\end{subequations}
Notably, both $\tilde{C}_{d2}$ and $\tilde{C}_{dc}$ remain nonzero even in the complete absence of the corresponding direct parasitic capacitances, i.e., when $C_{d2u,1} = C_{d2d,1} = 0$ and $C_{dc,1} = 0$. 
 
The induced terms $-\mleft[ C_{+-}^T C_{++}^{-1} q_{d,+} \mright]_{2-}$ and $-\mleft[ C_{+-}^T C_{++}^{-1} q_{d,+} \mright]_{c}$ alone generate an effective drive on Q$_2$ and on the coupler, respectively, since eliminating the common-mode sector mixes all three differential-mode charges together. A key consequence is that, in $q_{d,\rm eff}$, the drive $V_{d1}$ on one floating qubit couples to both the coupler and the other qubit even when no direct stray capacitance connects them. This is equivalent to three independent, simultaneous drives acting on each node flux in a circuit with two grounded qubits and a grounded coupler. The same argument holds for the case of the XY$_2$  control line ($V_{d1} = 0$) such that the drive on Q$_2$ gives non-zero coupling to Q$_1$ and the coupler, even if $C_{d1u,2} = C_{d1d,2} = C_{dc,2} = 0$.

The effective Hamiltonian for the differential sector now follows from the Legendre transform restricted to $(q_-, \dot\phi_-)$, where $\mathcal{L}_{\rm eff}$ is $\mathcal{L}$ with $\dot\phi_+$ eliminated:
\begin{align}
  \mathcal{H}_{\rm eff} =&\ q_-^T \dot{\phi}_- - \mathcal{L}_{\rm eff} \nonumber\\
  =&\ \frac{1}{2} q_-^T C_{\rm eff}^{-1} q_- + q_-^T C_{\rm eff}^{-1} q_{d,\rm eff} \nonumber\\
  &- \sum_{k \in \{1,2,c\}} E_{Jk} \cos \mleft( \frac{(\phi_-)_k}{\varphi_0} \mright) \, .
\end{align}
%


\subsection{Quantization}

In order to quantize the system, we promote the flux and charge variables to operators. Since the system can be understood as three coupled weakly anharmonic oscillators, we express the operators in terms of creation and annihilation operators $a_k^\dag$ and $a_k$ for $k \in \{1,2,c\}$, writing the Hamiltonian as a set of coupled Kerr oscillators. Defining $X_k \equiv a_k^\dag - a_k$ for $k \in \{1,2,c\}$, we obtain
\begin{align}
  \mathcal{H} =&\ \sum_{k \in \{1,2,c\}} \omega_k a_k^\dag a_k \nonumber \\
  &+ \sum_{k \in \{1,2,c\}} \frac{\alpha_k}{2} \mleft[ a_k^{\dag 2} a_k^2 + \mleft( a_k^{\dag 2} + \tfrac{2}{3} a_k^{\dag 3} a_k + \tfrac{1}{6} a_k^{\dag 4} + \text{H.c.} \mright) \mright] \nonumber \\
  &+ \sum_{k \in \{1,2,c\}} i \epsilon_k X_k
  + \sum_{ij \in \{1c,2c,12\}} g_{ij} X_i X_j \, , 
  \label{eq:kerr_h_app}
\end{align}
where
\begin{align}
  \omega_{k} =&\ \sqrt{8 E_{c,kk} E_{Jk}} - E_{c,kk} \, , \\
  \alpha_{k} =&\ - E_{c,kk} \, , \qquad k \in \{1, 2, c\} \, , \\
  g_{ij} =&\ - E_{c,ij} \sqrt[4]{\frac{4 E_{Ji} E_{Jj}}{E_{c,ii} E_{c,jj}}} \, , \qquad ij \in \{12, 1c, 2c\} \, .\\
  \epsilon_1 =&\ \Big[\frac{V_{d1}}{e}\mleft(E_{c,11}\tilde{C}_{d1} + E_{c,12}\tilde{C}_{d2} + E_{c,1c}\tilde{C}_{dc}\mright) \nonumber \\
  &+ \frac{V_{d2}}{e}\mleft(E_{c,11}\breve{C}_{d1} + E_{c,12}\breve{C}_{d2} + E_{c,1c}\breve{C}_{dc}\mright)\Big] \sqrt[4]{\frac{8 E_{J1}}{E_{c,11}}} \, , \\
  \epsilon_2 =&\ \Big[\frac{V_{d1}}{e}\mleft(E_{c,12}\tilde{C}_{d1} + E_{c,22}\tilde{C}_{d2} + E_{c,2c}\tilde{C}_{dc}\mright) \nonumber \\
  &+ \frac{V_{d2}}{e}\mleft(E_{c,12}\breve{C}_{d1} + E_{c,22}\breve{C}_{d2} + E_{c,2c}\breve{C}_{dc}\mright)\Big] \sqrt[4]{\frac{8 E_{J2}}{E_{c,22}}} \, , \\
  \epsilon_c =&\ \Big[\frac{V_{d1}}{e}\mleft(E_{c,1c}\tilde{C}_{d1} + E_{c,2c}\tilde{C}_{d2} + E_{c,cc}\tilde{C}_{dc}\mright) \nonumber \\
  &+ \frac{V_{d2}}{e}\mleft(E_{c,1c}\breve{C}_{d1} + E_{c,2c}\breve{C}_{d2} + E_{c,cc}\breve{C}_{dc}\mright)\Big] \sqrt[4]{\frac{8 E_{Jc}}{E_{c,cc}}} \, .
\end{align}
%


\section{Grounded qubits}
\label{app:grounded_qubit}

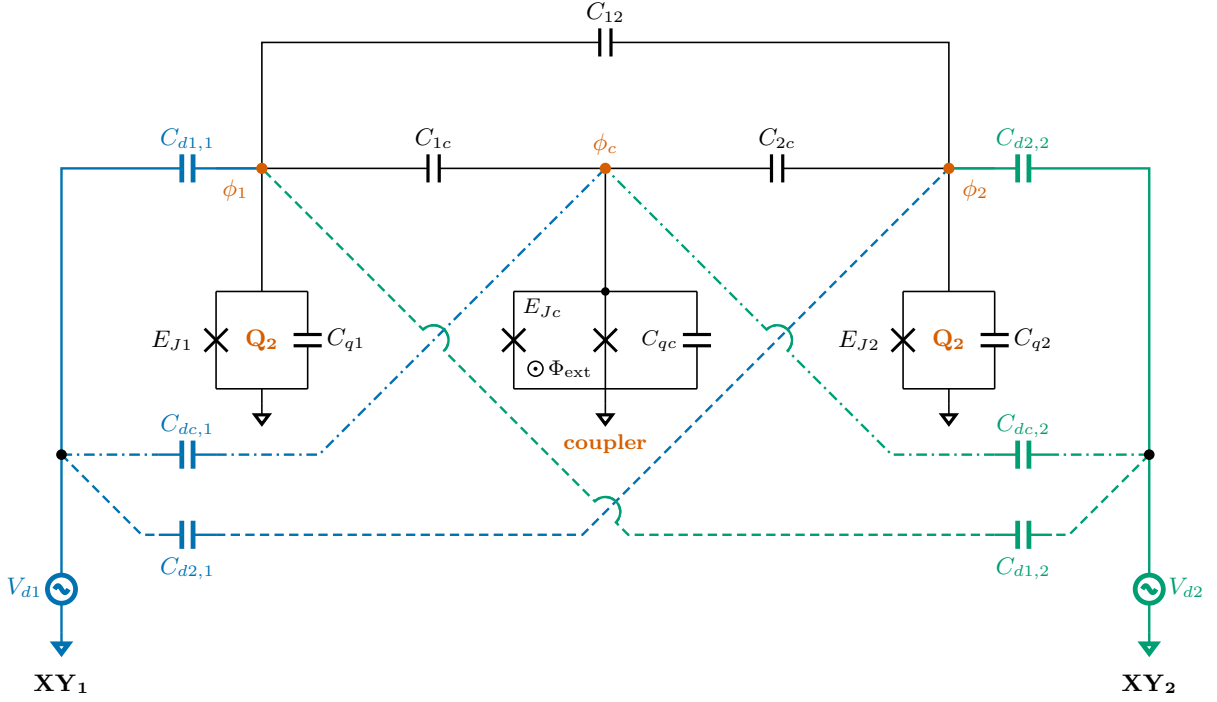
\begin{figure*}
    \centering
    \newcommand{\schematicwidth}{0.9\textwidth}
%
\begingroup
\ifdefined\schematicunit\else
  \ifdefined\schematicwidth
    \pgfmathsetlengthmacro{\schematicunit}{(\schematicwidth)/21.3}
  \else
    \def\schematicunit{0.68cm}
  \fi
\fi

\definecolor{xyone}{RGB}{0,114,178}    
\definecolor{xytwo}{RGB}{0,158,115}    
\definecolor{nodered}{RGB}{213,94,0}   

\tikzset{
  xy1/.style  = {color=xyone, line width=0.9pt},
  xy2/.style  = {color=xytwo, line width=0.9pt},
  par1/.style = {color=xyone, line width=0.9pt, dash pattern=on 3.5pt off 2pt},
  par2/.style = {color=xytwo, line width=0.9pt, dash pattern=on 3.5pt off 2pt},
  parc1/.style = {color=xyone, line width=0.9pt, dash pattern=on 4pt off 1.8pt on 0.9pt off 1.8pt},
  parc2/.style = {color=xytwo, line width=0.9pt, dash pattern=on 4pt off 1.8pt on 0.9pt off 1.8pt},
  phinode/.style  = {circle, fill=nodered, inner sep=1.5pt},
  philabel/.style = {text=nodered},
}
\ctikzset{bipoles/length=0.62cm, capacitors/width=0.24, capacitors/height=0.6}

\providecommand{\JJ}[1]{%
  \draw[line width=1.1pt] ($#1+(-0.2,-0.2)$) -- ($#1+(0.2,0.2)$)
                          ($#1+(-0.2,0.2)$) -- ($#1+(0.2,-0.2)$);}

\begin{circuitikz}[x=\schematicunit, y=\schematicunit, line width=0.6pt, font=\small]
\def\yn{6}

\draw (3.2,\yn) -- (6.3,\yn) to[C, l=$C_{1c}$] (7.7,\yn) -- (12.3,\yn)
      to[C, l=$C_{2c}$] (13.7,\yn) -- (16.8,\yn);

\draw (4,\yn) -- (4,8.2) -- (9.3,8.2) to[C, l=$C_{12}$] (10.7,8.2) -- (16,8.2) -- (16,\yn);

\draw (4,\yn) -- (4,3.85);
\draw (3.2,3.85) -- (4.8,3.85)  (3.2,2.15) -- (4.8,2.15);
\draw (3.2,3.85) -- (3.2,2.15);  \JJ{(3.2,3)}
\node[anchor=east] at (2.95,3) {$E_{J1}$};
\draw (4.8,3.85) to[C, l=$C_{q1}$] (4.8,2.15);
\node[text=nodered] at (4,3) {$\mathbf{Q}_{\mathbf{2}}$};
\draw (4,2.15) -- (4,1.9) node[sground]{};

\draw (16,\yn) -- (16,3.85);
\draw (15.2,3.85) -- (16.8,3.85)  (15.2,2.15) -- (16.8,2.15);
\draw (15.2,3.85) -- (15.2,2.15);  \JJ{(15.2,3)}
\node[anchor=east] at (14.95,3) {$E_{J2}$};
\draw (16.8,3.85) to[C, l=$C_{q2}$] (16.8,2.15);
\node[text=nodered] at (16,3) {$\mathbf{Q}_{\mathbf{2}}$};
\draw (16,2.15) -- (16,1.9) node[sground]{};

\draw (10,\yn) -- (10,3.85);
\draw (8.4,3.85) -- (11.6,3.85)  (8.4,2.15) -- (11.6,2.15);
\draw (8.4,3.85) -- (8.4,2.15);   \JJ{(8.4,3)}
\draw (10,3.85)  -- (10,2.15);    \JJ{(10,3)}
\draw (11.6,3.85) to[C, l_={\footnotesize $C_{qc}$}] (11.6,2.15);
\node[anchor=north west, font=\footnotesize, inner sep=1pt] at (8.5,3.78) {$E_{Jc}$};
\draw[line width=0.6pt] (8.78,2.5) circle (0.13);  \fill (8.78,2.5) circle (0.035);
\node[anchor=west, font=\footnotesize, inner sep=1pt] at (8.94,2.5) {$\Phi_{\mathrm{ext}}$};
\draw (10,2.15) -- (10,1.9) node[sground]{};
\node[text=nodered, font=\footnotesize\bfseries] at (10,1.2) {coupler};

\coordinate (H1) at (0.5,1);
\draw[xy1] (H1) -- (0.5,\yn) -- (2.2,\yn) to[C, l=$C_{d1,1}$] (3.2,\yn) -- (4,\yn);
\draw[parc1] (H1) -- (2.2,1);   \draw[xy1] (2.2,1) to[C, l=$C_{dc,1}$] (3.2,1);
\draw[par1]  (H1) -- (1.9,-0.4) -- (2.2,-0.4);
\draw[xy1] (2.2,-0.4) to[C, l_=$C_{d2,1}$] (3.2,-0.4);
\draw[xy1] (H1) -- (0.5,-0.6) to[sV, l_=$V_{d1}$] (0.5,-2.1) node[sground]{};
\fill (H1) circle (2pt);
\node at (0.5,-3) {$\mathbf{XY}_{\mathbf{1}}$};

\coordinate (H2) at (19.5,1);
\draw[xy2] (H2) -- (19.5,\yn) -- (17.8,\yn) to[C, l_=$C_{d2,2}$] (16.8,\yn) -- (16,\yn);
\draw[parc2] (H2) -- (17.8,1);  \draw[xy2] (17.8,1) to[C, l_=$C_{dc,2}$] (16.8,1);
\draw[par2]  (H2) -- (18.1,-0.4) -- (17.8,-0.4);
\draw[xy2] (17.8,-0.4) to[C, l=$C_{d1,2}$] (16.8,-0.4);
\draw[xy2] (H2) -- (19.5,-0.6) to[sV, l=$V_{d2}$] (19.5,-2.1) node[sground]{};
\fill (H2) circle (2pt);
\node at (19.5,-3) {$\mathbf{XY}_{\mathbf{2}}$};
\draw[parc1] (3.2,1) -- (5,1) -- (10,\yn);          
\draw[par1]  (3.2,-0.4) -- (9.6,-0.4) -- (16,\yn);  
\def\hr{0.26}
\draw[par2] (16.8,-0.4) -- (10.4,-0.4)
    -- ($(10,0)+(-45:\hr)$) arc[start angle=-45, end angle=135, radius=\hr]
    -- ($(7,3)+(-45:\hr)$)  arc[start angle=-45, end angle=135, radius=\hr]
    -- (4,\yn);                                       
\draw[parc2] (16.8,1) -- (15,1)
    -- ($(13,3)+(-45:\hr)$) arc[start angle=-45, end angle=135, radius=\hr]
    -- (10,\yn);                                      
\begin{scope}[color=xytwo, line width=0.9pt]
  \draw ($(10,0)+(-45:\hr)$) arc[start angle=-45, end angle=135, radius=\hr];
  \draw ($(7,3)+(-45:\hr)$)  arc[start angle=-45, end angle=135, radius=\hr];
  \draw ($(13,3)+(-45:\hr)$) arc[start angle=-45, end angle=135, radius=\hr];
\end{scope}

\fill (10,3.85) circle (1.6pt);
\node[phinode] at (4,\yn) {};   \node[philabel, anchor=north east] at (3.92,5.97) {$\phi_{1}$};
\node[phinode] at (10,\yn) {};  \node[philabel, anchor=south] at (10,6.08) {$\phi_{c}$};
\node[phinode] at (16,\yn) {};  \node[philabel, anchor=north west] at (16.08,5.97) {$\phi_{2}$};

\end{circuitikz}
\endgroup
\caption{Schematic of the grounded-qubit circuit, consisting of two grounded transmon qubits, Q$_1$ and Q$_2$, capacitively coupled to a grounded flux-tunable coupler through $C_{1c}$ and $C_{2c}$, with direct qubit-qubit coupling through $C_{12}$. The XY$_1$ (blue) and XY$_2$ (green) drive lines couple capacitively to the circuit. Solid lines indicate coupling to the target qubit, dash-dotted lines coupling to the coupler, and dashed lines coupling to the victim qubit. The corresponding drive capacitances are labeled in the schematic.}
    \label{fig:circuit_schematic_grounded}
\end{figure*}

We now derive the Hamiltonian for grounded qubits instead of floating ones. The grounded-transmon circuit we consider is shown in \figref{fig:circuit_schematic_grounded}. Following the same procedure as for floating qubits in \secref{sec:model} and \appref{app:LagrangianToHamiltonian}, we first write the Lagrangian in the node-flux basis, $\mathbf{\phi}^T = (\phi_1, \phi_2, \phi_c)$, as
\begin{align}
  \mathcal{L}
  =&\ \frac{1}{2} \mathbf{\dot{\phi}}^T C
     \mathbf{\dot{\phi}} \nonumber -\ \mathbf{\dot{\phi}}^T q_d
  + E_{J1} \cos \mleft( \frac{\phi_{1}}{\varphi_0} \mright) \\
  &+ E_{J2} \cos \mleft( \frac{\phi_{2} }{\varphi_0} \mright) + E_{Jc}(\Phi_{\text{ext}}) \cos \mleft( \frac{\phi_c}{\varphi_0} \mright) \, ,
\end{align}
where $C$ is the capacitance matrix of the network,
\begin{equation}
C =
\begin{pmatrix}
C_{\Sigma 1} & -C_{12} & -C_{1c} \\
-C_{12} & C_{\Sigma 2} & -C_{2c} \\
-C_{1c} & -C_{2c} & C_{\Sigma c}
\end{pmatrix} \, ,
\label{eq:C_grounded_matrix}
\end{equation}
with the diagonal elements representing the effective node capacitances:
\begin{align}
C_{\Sigma 1} &= C_{q1} + C_{12} + C_{1c} + C_{d1,1} + C_{d1,2} \, , \nonumber \\
C_{\Sigma 2} &= C_{q2} + C_{12} + C_{2c} + C_{d2,2} + C_{d2,1} \, , \nonumber \\
C_{\Sigma c} &= C_{qc} + C_{1c} + C_{2c} + C_{dc,1} + C_{dc,2} \, .
\end{align}
The charge-drive vector $q_d$ is given by
\begin{equation}
  q_d = V_{d1}
  \begin{pmatrix}
    C_{d1,1} \\
    C_{d2,1} \\
    C_{dc,1}
  \end{pmatrix} + V_{d2}
  \begin{pmatrix}
    C_{d1,2} \\
    C_{d2,2} \\
    C_{dc,2} 
  \end{pmatrix} \, .
\end{equation}
In this case, contrary to floating-qubit case, if the stray capacitances $C_{d2,1}$ and $C_{dc,1}$ become zero, then there is no induced drive term on Q$_2$ and the coupler node from XY$_1$.

We can obtain the conjugate charges
\begin{equation}
q = \frac{\partial \mathcal{L}}{\partial \dot\phi} = C \phi - q_d \Rightarrow \dot\phi = C^{-1}(q + q_d) \, . 
\end{equation}
The effective Hamiltonian then follows:
\begin{align}
  \mathcal{H} &= q^T \mathbf{\dot{\phi}} - \mathcal{L} \nonumber\\
  &=\ \frac{1}{2} (q+q_d)^T C^{-1}(q+q_d) - \sum_{k \in \{1,2,c\}} E_{Jk} \cos \mleft( \frac{\phi_k}{\varphi_0} \mright)  \,.
\end{align}

Now, in order to quantize the system, we promote the flux and charge variables to operators. Since the system can be understood as three coupled weakly anharmonic oscillators, we express the operators in terms of creation and annihilation operators. Defining $X_k \equiv a_k^\dag - a_k$ for $k \in \{1,2,c\}$, we obtain
\begin{align}
  \mathcal{H} =&\ \sum_{k \in \{1,2,c\}} \omega_k a_k^\dag a_k \nonumber \\
  &+ \sum_{k \in \{1,2,c\}} \frac{\alpha_k}{2} \mleft[ a_k^{\dag 2} a_k^2
  + \mleft( a_k^{\dag 2} + \tfrac{2}{3} a_k^{\dag 3} a_k + \tfrac{1}{6} a_k^{\dag 4} + \text{H.c.} \mright) \mright] \nonumber \\
  &+ \sum_{k \in \{1,2,c\}} i \epsilon_k X_k
  + g_{12} X_1 X_2 + \mleft(g_{1c} X_1 + g_{2c} X_2 \mright) X_c \, , 
  \label{eq:kerr_h_grounded}
\end{align}
where
\begin{align}
  \omega_{k} =&\ \sqrt{8 E_{c,kk} E_{Jk}} - E_{c,kk} \, , \\
  \alpha_{k} =&\ - E_{c,kk} \, , \qquad k \in \{1, 2, c\} \, , \\
  g_{ij} =&\ - E_{c,ij} \sqrt[4]{\frac{4 E_{Ji} E_{Jj}}{E_{c,ii} E_{c,jj}}} \, , \qquad ij \in \{12, 1c, 2c\} \, ,\\
  \epsilon_1 =&\ \sqrt[4]{\frac{8 E_{J1}}{E_{c,11}}} \mleft[\frac{V_{d1}}{e}\mleft(E_{c,11}{C}_{d1,1} + E_{c,12}{C}_{d2,1} + E_{c,1c}{C}_{dc,1}\mright) \mright. \nonumber \\
  &+ \mleft. \frac{V_{d2}}{e}\mleft(E_{c,11}{C}_{d1,2} + E_{c,12}{C}_{d2,2} + E_{c,1c}{C}_{dc,2}\mright) \mright] \, , \\
  \epsilon_2 =&\ \sqrt[4]{\frac{8 E_{J2}}{E_{c,22}}} \mleft[ \frac{V_{d1}}{e}\mleft(E_{c,12}{C}_{d1,1} + E_{c,22}{C}_{d2,1} + E_{c,2c}{C}_{dc,1} \mright) \mright. \nonumber \\
  &+ \mleft. \frac{V_{d2}}{e}\mleft(E_{c,12}{C}_{d1,2} + E_{c,22}{C}_{d2,2} + E_{c,2c}{C}_{dc,2}\mright) \mright] \, , \\
  \epsilon_c =&\ \sqrt[4]{\frac{8 E_{Jc}}{E_{c,cc}}} \mleft[ \frac{V_{d1}}{e}\mleft(E_{c,1c}{C}_{d1,1} + E_{c,2c}{C}_{d2,1} + E_{c,cc}{C}_{dc,1} \mright) \mright. \nonumber \\
  &+ \mleft. \frac{V_{d2}}{e}\mleft(E_{c,1c}{C}_{d1,2} + E_{c,2c}{C}_{d2,2} + E_{c,cc}{C}_{dc,2}\mright) \mright] \, .
\end{align}
For this grounded-qubit configuration, the remaining steps, namely transforming to the rotating frame, applying the high-frequency expansion, and diagonalizing into dressed modes, proceed identically to \secref{sec:model} in the main text (with details in \appref{app:magnus} and \appref{app:diagonalization}) and are not repeated here.

\begin{table}
\centering
\caption{System parameters for an example grounded-transmon circuit, including the two qubits Q$_1$ and Q$_2$, the tunable coupler C ($\Phi_{\mathrm{ext}}=0$), their mutual coupling parameters, and the drive capacitances.}
\label{tab:ground_system_params}

\renewcommand{\arraystretch}{1.1}
\resizebox{\columnwidth}{!}{%
\begin{tabular}{lccc}
\hline\hline
\textbf{Single-node parameters} & Q$_1$ & Q$_2$ & C \\
\hline
Frequency, $\omega_k / 2\pi$ (GHz)         & $5.287$  & $5.198$  & $7.318$ \\
Anharmonicity, $\alpha_k / 2\pi$ (GHz)     & $-0.266$ & $-0.267$ & $-0.155$ \\
Josephson energy, $E_{Jk} / h$ (GHz)       & $14.5$ & $14.0$ & $45.0$ \\
Charging energy, $E_{C,kk} / h$ (GHz)      & $0.266$  & $0.267$  & $0.155$ \\
Shunt capacitance, $C_{qk}$ (fF)           & $70.0$   & $70.0$   & $120.0$ \\
\hline
\textbf{Coupling parameters} & Q$_1$-Q$_2$ & Q$_1$-C & Q$_2$-C \\
\hline
Coupling strength, $g_{ij} / 2\pi$ (GHz)   & $0.010$ & $0.085$ & $0.084$ \\
Cross-charging, $E_{C,ij} / h$ (GHz)       & $0.0010$ & $0.0053$ & $0.0053$ \\
Coupling capacitance, $C_{ij}$ (fF)        & $0.210$  & $2.5$    & $2.5$ \\
\hline
\textbf{Drive capacitances} & Q$_1$ & Q$_2$ & C \\
\hline
XY$_1$, $C_{dk,1}$ (aF)            & $100.0$   & $5.0$    & $20.0$ \\
XY$_2$, $C_{dk,2}$ (aF)            & $5.0$    & $100.0$    & $20.0$ \\
\hline\hline
\end{tabular}%
}
\end{table}

\begin{figure}
    \centering
   \includegraphics[width=\columnwidth]{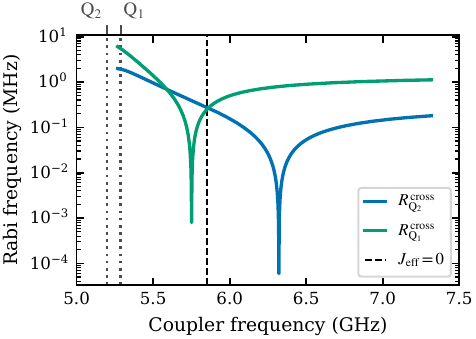}
\caption{Cross-Rabi frequencies for the grounded-qubit model, with XY$_1$ driving Q$_2$ and XY$_2$ driving Q$_1$ on resonance, calculated using the parameters in \tabref{tab:ground_system_params}. The dashed vertical line marks the coupler frequency at which $J_{\mathrm{eff}}=0$. The vertical dotted lines indicate the qubit frequencies. }
    \label{fig:grounded_rabi_cross}
\end{figure}

Using the representative parameter set listed in \tabref{tab:ground_system_params}, the resulting cross-Rabi minima for XY$_1$ driving Q$_2$ and XY$_2$ driving Q$_1$ on resonance are shown in \figref{fig:grounded_rabi_cross}.


\section{High-frequency expansion}
\label{app:magnus}

Here, we give the details for the derivation in \secref{sec:HFE} of the effective time-independent Hamiltonian in \eqref{eq:heff1_final} from a high-frequency expansion~\cite{Bukov2015, EckardtAnisimovas2015, goldman2014periodically} applied to \eqref{eq:DrivenHRotating}.

For a Hamiltonian periodic in time with drive frequency $\Omega$, i.e., satisfying $H'(t+2\pi/\Omega) = H'(t)$, the high-frequency expansion begins by first decomposing $H'(t)$ into a static component $H_0$ and its Fourier components:
\begin{equation}
  H' = H_0 + \sum_{j=1}^{\infty} \mleft[ V^{(j)}e^{ij\Omega t} + V^{(-j)}e^{-ij\Omega t} \mright] \, .
  \label{eq:app_harmonic_decomp}
\end{equation}
Here $V^{(-j)} = (V^{(j)})^\dag$ follows from the Hermiticity of $H'(t)$. Under the convergence condition $\Omega \gg H_0, V^{(j)}$, this decomposition allows a time-independent effective Hamiltonian to be constructed systematically as an expansion in $1/\Omega$~\cite{goldman2014periodically, Bukov2015, EckardtAnisimovas2015}:
\begin{align}
  H_{\rm eff} =&\ H_0 + \frac{1}{\Omega}\sum_{j=1}^{\infty} \frac{1}{j} \mleft[ V^{(j)}, V^{(-j)} \mright] \nonumber \\
  &+ \frac{1}{2\Omega^2} \sum_{j=1}^{\infty} \frac{1}{j^2} \mleft( \mleft[ \mleft[ V^{(j)}, H_0 \mright], V^{(-j)} \mright] + \text{H.c.} \mright) \nonumber \\
  &+ \frac{1}{3\Omega^2} \sum_{j,l=1}^{\infty} \frac{1}{jl} \mleft( \mleft[ V^{(j)}, \mleft[ V^{(l)}, V^{(-j-l)} \mright] \mright] \mright. \nonumber \\
  &\mleft. \qquad -\mleft[ V^{(j)}, \mleft[ V^{(-l)}, V^{(l-j)} \mright] \mright] + \text{H.c.} \mright) \nonumber \\
  &+ \mathcal{O}(1/\Omega^3) \, .
  \label{eq:app_hfe_general}
\end{align}
The first-order term ($1/\Omega$) involves pairwise commutators $[V^{(j)}, V^{(-j)}]$, and generates the leading renormalization of $H_0$'s own parameters. The two second-order terms ($1/\Omega^2$) involve, respectively, a nested commutator with $H_0$ and a triple commutator among three harmonics; both are systematically smaller by one further power of $1/\Omega$. The specific harmonic combinations retained at each order are fixed by the requirement that the resulting correction should be time-independent.

For the driven Hamiltonian in \eqref{eq:DrivenHRotating}, comparing \eqref{eq:app_harmonic_decomp} to the result in \eqref{eq:app_hfe_general} shows that only $j = 2$ contributes:
\begin{align}
  H_0 =&\ \sum_{k \in \{1,2,c\}} \mleft[ (\omega_k-\Omega) a_k^\dag a_k + \frac{\alpha_k}{2} a_k^\dag a_k^\dag a_k a_k \mright] \nonumber \\
  &- \sum_{ij \in \{1c,2c,12\}} g_{ij} \mleft[ a_i^\dag a_j + a_i a_j^\dag \mright] \nonumber \\
  &+ \frac{i\epsilon_1}{2} \sum_{k \in \{1,2,c\}} \lambda_{k1} \mleft( a_k^\dag -  a_k \mright) \, , \\
  V^{(2)} =&\ \sum_{ij \in \{1c,2c,12\}} g_{ij} a_i^\dag a_j^\dag + \frac{i\epsilon_1}{2} \sum_{k \in \{1,2,c\}} \lambda_{k1} a_k^\dag \, , \\
  V^{(-2)} =&\ \sum_{ij \in \{1c,2c,12\}} g_{ij} a_i a_j - \frac{i\epsilon_1}{2} \sum_{k \in \{1,2,c\}} \lambda_{k1}a_k \, .
\end{align}
Since the only nonzero $V$-terms are $V^{(\pm2)}$, the triple-commutator term in \eqref{eq:app_hfe_general} vanishes (for it to be non-zero would require a harmonic $V^{(0)}$ or $V^{(\pm4)}$, neither of which is present). The effective Hamiltonian therefore reduces, to $\mathcal{O}(1/\Omega^2)$, to
\begin{align}
  H_{\rm eff} =&\ H_0 + \frac{1}{2\Omega} \mleft[ V^{(2)} , V^{(-2)} \mright] \nonumber \\
  &+ \frac{1}{8\Omega^2} \mleft( \mleft[ \mleft[ V^{(2)} , H_0 \mright] , V^{(-2)} \mright] + \text{H.c.} \mright) \, .
  \label{eq:HeffFirstOrder_app}
\end{align}

Evaluating the first-order correction in \eqref{eq:HeffFirstOrder_app} (the leading order retained here) shows that it has the same structure as $H_0$ itself:
\begin{align}
  \frac{1}{2\Omega} \mleft[ V^{(2)} , V^{(-2)} \mright] =&\ -\sum_{k \in \{1,2,c\}} \delta \omega_k a_k^\dag a_k \nonumber \\
  &- \sum_{ij \in \{1c,2c,12\}} \delta g_{ij} \mleft( a_i^\dag a_j + a_i a_j^\dag \mright) \nonumber \\
  &+ \frac{i\epsilon_1}{2} \sum_{k \in \{1,2,c\}} \delta \lambda_{k1}\mleft( a_k^\dag -  a_k \mright) \nonumber \\
   & + \delta C \, ,
  \label{eq:V2Vm2_assembled}
\end{align}
where $\delta C$ is a constant energy offset and therefore does not affect the dynamics. Thus, adding this first-order correction to $H_0$ simply renormalizes each parameter as $\omega_k \to \tilde\omega_k \equiv \omega_k - \delta\omega_k$, $g_{ij} \to \tilde g_{ij} \equiv g_{ij} + \delta g_{ij}$, and $\lambda_{k1} \to \tilde\lambda_{k1} \equiv \lambda_{k1} + \delta\lambda_{k1}$. Explicitly, the renormalized parameters are
\begin{align}
  \tilde\omega_1 &= \omega_1 - \frac{g_{12}^2 + g_{1c}^2}{2\Omega} \, , \quad
  \tilde\omega_2 = \omega_2 - \frac{g_{12}^2 + g_{2c}^2}{2\Omega} \, , \nonumber \\
  \tilde\omega_c &= \omega_c - \frac{g_{1c}^2 + g_{2c}^2}{2\Omega} \, ,
  \label{eq:tilde_omega}
\\
  \tilde g_{12} &= g_{12} + \frac{g_{1c}g_{2c}}{2\Omega} \, , \quad
  \tilde g_{1c} = g_{1c} + \frac{g_{12}g_{2c}}{2\Omega} \, , \nonumber \\
  \tilde g_{2c} &= g_{2c} + \frac{g_{12}g_{1c}}{2\Omega} \, ,
  \label{eq:tilde_g}
\\
  \tilde\lambda_{11} &= \lambda_{11} + \frac{g_{12}\lambda_{21} + g_{1c}\lambda_{c1}}{2\Omega} \, , \quad
  \tilde\lambda_{21} = \lambda_{21} + \frac{g_{12}\lambda_{11} + g_{2c}\lambda_{c1}}{2\Omega} \, , \nonumber \\
  \tilde\lambda_{c1} &= \lambda_{c1} + \frac{g_{1c}\lambda_{11} + g_{2c}\lambda_{21}}{2\Omega} \, .
  \label{eq:tilde_Lambda}
\end{align}

The effective Hamiltonian in the rotating frame can therefore be written as
\begin{widetext}
\begin{align}
  H_{\rm eff}^{(1)} =&\ \mleft( \tilde\omega_1 - \Omega \mright) a_1^\dag a_1 + \frac{\alpha_1}{2} a_1^\dag a_1^\dag a_1 a_1 
  + \mleft( \tilde\omega_2 - \Omega \mright) a_2^\dag a_2 + \frac{\alpha_2}{2} a_2^\dag a_2^\dag a_2 a_2 
  + \mleft( \tilde\omega_c - \Omega \mright) a_c^\dag a_c + \frac{\alpha_c}{2} a_c^\dag a_c^\dag a_c a_c \nonumber \\
  &- \tilde g_{12} \mleft( a_1^\dag a_2 + a_1 a_2^\dag \mright) - \tilde g_{1c} \mleft( a_1^\dag a_c + a_1 a_c^\dag \mright) - \tilde g_{2c} \mleft( a_2^\dag a_c + a_2 a_c^\dag \mright) \nonumber \\
  &+ \frac{i\epsilon_1}{2} \tilde\lambda_{11} a_1^\dag + \frac{i\epsilon_1}{2} \tilde\lambda_{21} a_2^\dag + \frac{i\epsilon_1}{2} \tilde\lambda_{c1} a_c^\dag + \text{H.c.} 
  + \mathcal{O}(1/\Omega^2) \, ,
  \label{eq:heff1_final_app}
\end{align}
\end{widetext}
which is \eqref{eq:heff1_final} in the main text with all terms written out separately and explicitly without summation.


\section{Diagonalizing the effective Hamiltonian}
\label{app:diagonalization}

Here, we provide the details of the diagonalization performed in \secref{sec:dressing} to go from the effective time-independent Hamiltonian in \eqref{eq:heff1_final} to its normal-mode form in \eqref{eq:heff_dressed_final}. While this amounts to a standard Bogoliubov transformation, we show the whole procedure here for completeness and transparency about how all renormalized parameters are obtained.


\subsection{Setting up the diagonalization}

To find the unitary rotation from the bare modes $a_1$, $a_2$, and $a_c$ in \eqref{eq:heff1_final} to the normal modes $b_1$, $b_2$, and $b_c$ in \eqref{eq:heff_dressed_final}, we make the ansatz
\begin{align}
  a_1 =&\ \gamma_{11}b_1 + \gamma_{12}b_2 + \gamma_{1c}b_c \, , \label{eq:ansatz_a1} \\
  a_2 =&\ \gamma_{21}b_1 + \gamma_{22}b_2 + \gamma_{2c}b_c \, , \label{eq:ansatz_a2} \\
  a_c =&\ \gamma_{c1}b_1 + \gamma_{c2}b_2 + \gamma_{cc}b_c \, . \label{eq:ansatz_ac}
\end{align}
Here, $\gamma_{km}$ denotes the matrix element connecting bare mode
$k\in\{1,2,c\}$ to normal mode $m\in\{1,2,c\}$. Grouping these
coefficients by column defines
$\gamma_{1} \equiv (\gamma_{11}, \gamma_{21}, \gamma_{c1})^T$,
$\gamma_{2} \equiv (\gamma_{12}, \gamma_{22}, \gamma_{c2})^T$, and
$\gamma_{c} \equiv (\gamma_{1c}, \gamma_{2c}, \gamma_{cc})^T$. Preserving the bosonic algebra for $b_1$, $b_2$, and $b_c$ requires $\gamma$ to be unitary.

The quadratic part of $H_{\rm eff}^{(1)}$ in \eqref{eq:heff1_final}, excluding the quartic anharmonic term $\frac{\alpha_k}{2} a_k^\dag a_k^\dag a_k a_k$, is
\begin{align}
  H_{\rm quad} =&\ \sum_{k\in\{1,2,c\}}(\tilde\omega_k-\Omega)\,a_k^\dag a_k \nonumber \\
  &- \sum_{ij \in \{1c,2c,12\}} \tilde g_{ij} \mleft(a_i^\dag a_j + a_i a_j^\dag \mright) \, .
  \label{eq:Hquad}
\end{align}
Equation~(\ref{eq:Hquad}) can be written compactly as $H_{\rm quad} = \sum_{k,l} M_{kl} a_k^\dag a_l$, with $M$ the real symmetric matrix
\begin{equation}
  M = \begin{pmatrix}
    \tilde\omega_1 - \Omega & -\tilde g_{12} & -\tilde g_{1c} \\
    -\tilde g_{12} & \tilde\omega_2 - \Omega & -\tilde g_{2c} \\
    -\tilde g_{1c} & -\tilde g_{2c} & \tilde\omega_c - \Omega
  \end{pmatrix} \, ,
  \label{eq:app_M}
\end{equation}
whose diagonal entries are the renormalized single-mode detunings and whose off-diagonal entries are the renormalized pairwise couplings. Since $M$ is real and symmetric, it is diagonalized by a unitary $\gamma$ whose entries $\gamma_{km}$ can be taken real, such that $\gamma^\dag M \gamma = \mathrm{diag}(\omega_{\rm eff,1} - \Omega, \omega_{\rm eff,2} - \Omega, \omega_{\rm eff,c} - \Omega)$, where $\omega_{{\rm eff},m}$ are the lab-frame dressed-mode frequencies of the qubit-1-like, qubit-2-like, and coupler-like modes.


\subsection{Dressed-mode frequencies}

To find the dressed-mode frequencies, we solve for the eigenvalues of $M$. These eigenvalues satisfy $\det \mleft[ M - (\omega_{\rm eff} - \Omega) I \mright] = 0$, which expands to the cubic equation
\begin{equation}
  (\omega_{\rm eff} - \Omega)^3 - I_1(\omega_{\rm eff} - \Omega)^2 + I_2(\omega_{\rm eff} - \Omega) - I_3 = 0 \, ,
  \label{eq:app_charpoly}
\end{equation}
with the matrix invariants
\begin{align}
  I_1 =&\ \mathrm{Tr}(M) = \tilde\omega_1 + \tilde\omega_2 + \tilde\omega_c - 3\Omega \, , \\
  I_2 =&\ \frac{1}{2} \mleft[ \mathrm{Tr}(M)^2 - \mathrm{Tr}(M^2) \mright] \, , \\
  I_3 =&\ \det M \, .
\end{align}

Since $M$ is real symmetric, all three roots of \eqref{eq:app_charpoly} are real. Substituting $\omega_{\rm eff} - \Omega = \mu + I_1/3$ removes the quadratic term, giving the depressed cubic $\mu^3 + p\mu + q = 0$ with~\cite{Siddique2020, janson2010roots}
\begin{equation}
  p = I_2 - \frac{I_1^2}{3} \, , \qquad q = -\frac{2I_1^3}{27} + \frac{I_1 I_2}{3} - I_3 \, .
\end{equation}
Evaluating these invariants explicitly for $M$ and simplifying gives
\begin{widetext}
\begin{align}
  p =&\ -\frac{1}{6} \mleft[(\tilde\omega_1 - \tilde\omega_2)^2 + (\tilde\omega_2 - \tilde\omega_c)^2 + (\tilde\omega_c - \tilde\omega_1)^2 \mright] - \mleft(\tilde g_{12}^2 + \tilde g_{1c}^2 + \tilde g_{2c}^2\mright) \, ,
  \label{eq:app_p} \\
  q =&\ -\frac{2}{27} \mleft( \tilde\omega_1^3 + \tilde\omega_2^3 + \tilde\omega_c^3 \mright) 
  + \frac{1}{9} \mleft( \tilde\omega_1^2 \tilde\omega_2 + \tilde\omega_1^2 \tilde\omega_c + \tilde\omega_2^2 \tilde\omega_c \mright) 
  + \frac{1}{9} \mleft( \tilde\omega_1 \tilde\omega_2^2 + \tilde\omega_1 \tilde\omega_c^2 + \tilde\omega_2 \tilde\omega_c^2 \mright) - \frac{4}{9} \tilde\omega_1 \tilde\omega_2 \tilde\omega_c \nonumber \\
  &- \frac{1}{3} \mleft[ \tilde\omega_1 \mleft( \tilde g_{12}^2 + \tilde g_{1c}^2 - 2\tilde g_{2c}^2 \mright)
  + \tilde\omega_2 \mleft( \tilde g_{12}^2 - 2\tilde g_{1c}^2 + \tilde g_{2c}^2 \mright)
  + \tilde\omega_c \mleft( -2\tilde g_{12}^2 + \tilde g_{1c}^2 + \tilde g_{2c}^2 \mright) \mright] 
  + 2 \tilde g_{12} \tilde g_{1c} \tilde g_{2c} \, .
  \label{eq:app_q}
\end{align}
\end{widetext}
Notably, $p$ and $q$ only depend on differences and products of $\tilde\omega_1$, $\tilde\omega_2$, $\tilde\omega_c$, $\tilde g_{12}$, $\tilde g_{1c}$, and $\tilde g_{2c}$ as the explicit $\Omega$-dependence in $I_1$, $I_2$, and $I_3$ cancels entirely. Since $p < 0$ and all three roots are real, the depressed cubic falls in the trigonometric regime, with solutions $\mu_n = 2\sqrt{-p/3} \cos(\theta - 2\pi n/3)$, $n = 0, 1, 2$, and
\begin{equation}
  \theta = \frac{1}{3}\arccos \mleft( \frac{3q}{2p} \sqrt{-\frac{3}{p}} \mright) \, .
  \label{eq:app_theta}
\end{equation}
Substituting back, $\omega_{\rm eff}^{(n)} = \mu_n + (\tilde\omega_1 + \tilde\omega_2 + \tilde\omega_c)/3$, where the $\Omega$ in $I_1$ cancels, gives the three roots
\begin{equation}
  \omega_{\rm eff}^{(n)} = \frac{\tilde\omega_1 + \tilde\omega_2 + \tilde\omega_c}{3} + 2\sqrt{-\frac{p}{3}} \cos \mleft( \theta - \frac{2\pi n}{3} \mright) \, , \quad n = 0, 1, 2 \, ,
  \label{eq:eigenvalues}
\end{equation}
with $p, q, \theta$ from Eqs.~(\ref{eq:app_p})--(\ref{eq:app_theta}). The index $n$ orders the roots by magnitude, with $n=0$ the highest and $n=2$ the lowest, and not by physical mode. The effective mode frequencies $\omega_{{\rm eff},1}$, $\omega_{{\rm eff},2}$, and $\omega_{{\rm eff},c}$ of the qubit-1-like, qubit-2-like, and coupler-like modes are obtained by assigning each root $\omega_{\rm eff}^{(n)}$ to the bare mode with which its eigenvector has the largest overlap, as described in \secref{app:gamma}.


\subsection{Mode-rotation matrix}
\label{app:gamma}

Next, we determine the mode-rotation matrix $\gamma$. Since $\gamma^\dag M \gamma$ is diagonal, the columns of $\gamma$ are the normalized eigenvectors of $M$. For each root $\omega_{\rm eff}^{(n)}$ from \eqref{eq:eigenvalues}, the corresponding unnormalized eigenvector $v^{(n)}$ is a nonzero solution of
\begin{equation}
  \mleft( M - \mleft( \omega_{\rm eff}^{(n)} - \Omega \mright) I \mright) v^{(n)} = 0 \, .
\end{equation}
Since this matrix is singular, the cross product of any two of its rows gives an eigenvector. Taking rows 1 and 2, this gives
\begin{equation}
  v^{(n)} = \begin{pmatrix}
    \tilde g_{1c} \mleft( \tilde\omega_2-\omega_{\rm eff}^{(n)} \mright) + \tilde g_{12}\tilde g_{2c} \\
    \tilde g_{2c} \mleft( \tilde\omega_1-\omega_{\rm eff}^{(n)} \mright) + \tilde g_{12}\tilde g_{1c} \\
    \mleft( \tilde\omega_1 - \omega_{\rm eff}^{(n)} \mright) \mleft( \tilde\omega_2 - \omega_{\rm eff}^{(n)} \mright) - \tilde g_{12}^2
  \end{pmatrix} \, ,
  \label{eq:app_eigenvector}
\end{equation}
whose entries $(v^{(n)})_k$ correspond to the bare modes $k=1,2,c$, with norm
\begin{align}
  |v^{(n)}| =&\ \mleft[ \mleft( \tilde g_{1c} \mleft( \tilde\omega_2 - \omega_{\rm eff}^{(n)} \mright) + \tilde g_{12} \tilde g_{2c} \mright)^2 \mright. \nonumber \\
  &+ \mleft( \tilde g_{2c} \mleft( \tilde\omega_1 - \omega_{\rm eff}^{(n)} \mright) + \tilde g_{12} \tilde g_{1c} \mright)^2 \nonumber \\
  &+ \mleft. \mleft( \mleft( \tilde\omega_1 - \omega_{\rm eff}^{(n)} \mright) \mleft( \tilde\omega_2 - \omega_{\rm eff}^{(n)} \mright) - \tilde g_{12}^2 \mright)^2 \mright]^{1/2} \, .
  \label{eq:app_norm}
\end{align}

The three eigenpairs of $M$ are matched one-to-one with the bare modes $k \in \{1, 2, c\}$ by selecting the assignment that maximizes the sum of the bare-mode weights $|(v^{(n)})_k|^2/|v^{(n)}|^2$. This assignment fixes the mode labels $m \in \{1, 2, c\}$, with the corresponding eigenvalue defining $\omega_{{\rm eff},m}$ and the normalized eigenvector defining the column $\gamma_m$ of $\gamma$.

Explicitly, the three columns of $\gamma$ are
\begin{align}
  \gamma_{1} =&\ \frac{1}{|v_1|}\begin{pmatrix}
    \tilde g_{1c}\mleft(\tilde\omega_2-\omega_{{\rm eff},1}\mright) + \tilde g_{12}\tilde g_{2c} \\
    \tilde g_{2c}\mleft(\tilde\omega_1-\omega_{{\rm eff},1}\mright) + \tilde g_{12}\tilde g_{1c} \\
    \mleft(\tilde\omega_1-\omega_{{\rm eff},1}\mright)\mleft(\tilde\omega_2-\omega_{{\rm eff},1}\mright) - \tilde g_{12}^2
  \end{pmatrix} \, , \\
  \gamma_{2} =&\ \frac{1}{|v_2|}\begin{pmatrix}
    \tilde g_{1c} \mleft( \tilde\omega_2 - \omega_{{\rm eff},2} \mright) + \tilde g_{12} \tilde g_{2c} \\
    \tilde g_{2c} \mleft( \tilde\omega_1 - \omega_{{\rm eff},2} \mright) + \tilde g_{12} \tilde g_{1c} \\
    \mleft( \tilde\omega_1 - \omega_{{\rm eff},2} \mright) \mleft( \tilde\omega_2 - \omega_{{\rm eff},2} \mright) - \tilde g_{12}^2
  \end{pmatrix} \, , \\
  \gamma_{c} =&\ \frac{1}{|v_c|}\begin{pmatrix}
    \tilde g_{1c} \mleft( \tilde\omega_2 - \omega_{{\rm eff},c} \mright) + \tilde g_{12} \tilde g_{2c} \\
    \tilde g_{2c} \mleft( \tilde\omega_1 - \omega_{{\rm eff},c} \mright) + \tilde g_{12} \tilde g_{1c} \\
    \mleft( \tilde\omega_1 - \omega_{{\rm eff},c} \mright) \mleft( \tilde\omega_2 - \omega_{{\rm eff},c} \mright) - \tilde g_{12}^2
  \end{pmatrix} \, ,
  \label{eq:app_gamma_full}
\end{align}
where $|v_m|$ is the norm in \eqref{eq:app_norm} evaluated for the eigenvector assigned to mode $m$.


\subsection{The diagonalized Hamiltonian}

Using the dressed-mode frequencies $\omega_{{\rm eff},m}$ and the mode-rotation matrix $\gamma$, substituting the ansatz from Eqs.~(\ref{eq:ansatz_a1})--(\ref{eq:ansatz_ac}) into $H_{\rm quad}$ in \eqref{eq:Hquad} diagonalizes it exactly, giving
\begin{equation}
H_{\rm quad, d} = \sum_{m \in \{1,2,c\}} \mleft( \omega_{{\rm eff},m} - \Omega \mright) b_m^\dag b_m \, .
\end{equation}
Similarly, the drive term from \eqref{eq:heff1_final} becomes
\begin{equation}
 H_{\rm drive, d} = \frac{i\epsilon_1}{2} \sum_{m \in \{1,2,c\}} \Theta_m \mleft( b_m^\dag - b_m \mright) \, ,
\end{equation}
defining the dressed drive amplitude
\begin{equation}
  \Theta_m \equiv \sum_{k \in \{1,2,c\}} \gamma_{km} \tilde\lambda_{k1} \, .
  \label{eq:Theta_m}
\end{equation}
To leading order, the quartic term in~\eqref{eq:heff1_final} produces self-Kerr and cross-Kerr contributions such that the leading-order effective Hamiltonian becomes
\begin{align}
  H_{\rm eff, d}^{(1)} =&\ \sum_{m \in \{1,2,c\}} \mleft[ \mleft( \omega_{{\rm eff},m} - \Omega \mright) b_m^\dag b_m + \frac{\bar\alpha_m}{2} b_m^\dag b_m^\dag b_m b_m \mright] \nonumber \\
  &+ \sum_{mm' \in \{12,1c,2c\}} \chi_{mm'} b_m^\dag b_m b_{m'}^\dag b_{m'} \nonumber \\
  &+ \frac{i\epsilon_1}{2} \sum_{m \in \{1,2,c\}} \Theta_m \mleft( b_m^\dag - b_m \mright) \, ,
  \label{eq:heff_dressed_final_app}
\end{align}
which is \eqref{eq:heff_dressed_final} in the main text. Here, the self-Kerr and cross-Kerr coefficients are given by
\begin{align}
  \bar\alpha_m &\equiv \sum_{k \in \{1,2,c\}} \alpha_k|\gamma_{km}|^4 \, , \\
  \chi_{mm'} &\equiv 2 \sum_{k \in \{1,2,c\}} \alpha_k|\gamma_{km}|^2|\gamma_{km'}|^2 \, .
\end{align}
Substituting the explicit columns $\gamma_m$ from \eqref{eq:app_gamma_full} into the definition of $\Theta_m$ in \eqref{eq:Theta_m} gives its full expression in \eqref{eq:theta_explicit} in the main text.


\section{Comparison of drive-capacitance models}
\label{app:comparison_models}

\begin{figure}
    \centering
    \includegraphics[width=\columnwidth]{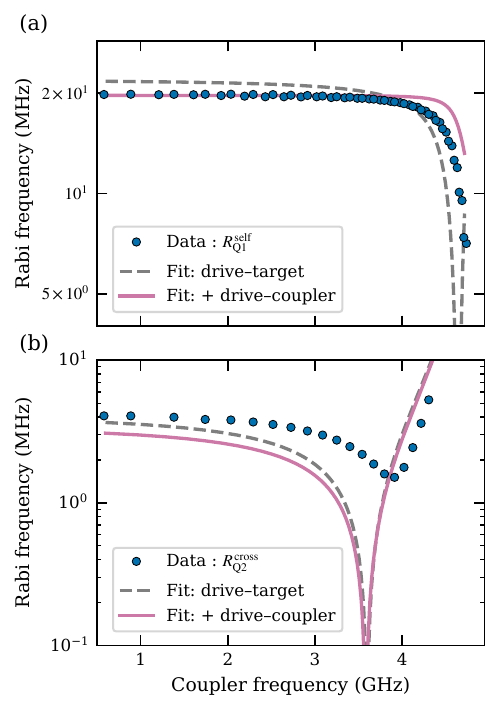}
\caption{Simultaneous fits to the measured (a) self-Rabi frequency of Q$_1$ and (b) cross-Rabi frequency of Q$_2$, using only the XY$_1$ target-qubit drive capacitances (dashed line) and with the drive-to-coupler capacitance additionally included (solid line). Neither model fits both datasets simultaneously.}
    \label{fig:XY1}
\end{figure}

To assess the role of the different drive-capacitance paths in reproducing the measured self- and cross-Rabi frequencies, we fit experimental data for XY$_1$ ($R^{\rm self}_{\rm Q1}$ and $R^{\rm cross}_{\rm Q2}$) to a hierarchy of models, as shown in \figref{fig:XY1}, using the drive-capacitance network defined in \figref{fig:circuit_schematic}. We first include only the XY$_1$ drive capacitances to the target qubit (dashed line), and then additionally include the drive-to-coupler capacitance (solid line), while keeping the direct drive capacitances to the victim qubit equal to zero. For each model, the self- and cross-Rabi data are fitted simultaneously using a common set of drive-capacitance parameters. The designed capacitance values are used as the initial parameters, and the corresponding fit ranges and fitted values are summarized in \tabref{tab:XY1_hierarchy}. For the model including the drive-to-coupler capacitance, all extracted capacitance values reach the boundaries of their allowed ranges. 

As seen in \figref{fig:XY1}, both models fail to simultaneously reproduce the measured self- and cross-Rabi frequencies. In contrast, when all drive capacitances, including those to the victim qubit, are included, good agreement with both measurements is obtained, as shown in \figref{fig:rabi_grid} of the main text. This result demonstrates that the drive paths to the target qubit, coupler, and victim qubit are all required to capture the observed crosstalk. The corresponding analysis for XY$_2$ leads to the same conclusion; we omit those results here, as they do not provide additional insights.

\begin{table}
    \centering
    \caption{Capacitance values from fitting $R^{\rm self}_{\rm Q1}$ and $R^{\rm cross}_{\rm Q2}$ in \figref{fig:XY1} with models considering only the XY$_1$ drive-target coupling and with the drive-to-coupler capacitance subsequently added. All capacitances are given in aF.}
    \label{tab:XY1_hierarchy}
    
    \renewcommand{\arraystretch}{1.1}
    \setlength{\tabcolsep}{4pt}
    \begin{tabular}{lccc}
        \hline\hline
        Parameter & Target only & Target + coupler & Fit range \\
        \hline
        $C_{d1u,1}$ & $206.7$ & $20.0$  & $[20,500]$ \\
        $C_{d1d,1}$ & $189.3$ & $500.0$ & $[20,500]$ \\
        $C_{dc,1}$  & --      & $5.0$   & $[5,40]$   \\
        \hline\hline
    \end{tabular}
\end{table}


\section{Effect of common-mode drive-to-victim capacitance on the crosstalk-null frequency}
\label{app:victim_sum_cap}

For the floating-transmon architecture, \figpanel{fig:xtalk_nulls}{c} of the main text shows the dependence of the crosstalk-null frequency on the differential drive-to-victim capacitance $\Delta C_{\mathrm{victim}}$. Here, we additionally consider the corresponding common-mode drive-to-victim capacitances,
\begin{align}
\Sigma C_{\mathrm{victim},1}
&= C_{d1u,2}+C_{d1d,2} \, , \\
\Sigma C_{\mathrm{victim},2}
&= C_{d2u,1}+C_{d2d,1} \, .
\end{align}
This additional dependence follows from Eqs.~(\ref{eq:effective_drive_cap_d1})--(\ref{eq:effective_drive_cap_dc}); the direct drive contribution is determined by the differential capacitance, whereas the induced contribution contains $q_{d,+}$ and therefore also depends on the common-mode drive capacitances.

\begin{figure}
    \centering
    \includegraphics[width=\columnwidth]
    {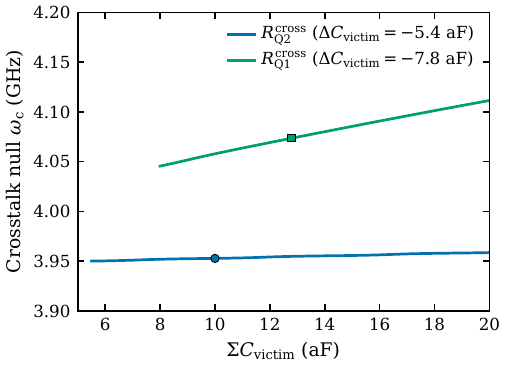}
\caption{Crosstalk-null frequencies for $R^{\rm cross}_{\rm Q2}$ (blue) and $R^{\rm cross}_{\rm Q1}$ (green) as a function of the corresponding common-mode drive-to-victim capacitance
$\Sigma C_{\mathrm{victim}}$, with
$\Delta C_{\mathrm{victim},2} = \SI{-5.4}{\atto\farad}$ and
$\Delta C_{\mathrm{victim},1} = \SI{-7.8}{\atto\farad}$, respectively, held fixed. 
The blue circle and green square indicate the $\Sigma C_{\mathrm{victim}}$ from the fitted capacitance values in \tabref{tab:drive_model_hierarchy}.}
    \label{fig:xtalk_null_C_sigma}
\end{figure}

To quantify this effect, we show the crosstalk-null frequencies as a function of $\Sigma C_{\mathrm{victim}}$ in \figref{fig:xtalk_null_C_sigma}, with the corresponding differential capacitances fixed at values obtained by the fitting in \figref{fig:rabi_grid}. We see in \figref{fig:xtalk_null_C_sigma} that the null frequencies vary only weakly with $\Sigma C_{\mathrm{victim}}$. Nevertheless, this dependence should be accounted for when choosing the drive capacitances to achieve a specific crosstalk-null frequency, since variations in $\Sigma C_{\mathrm{victim}}$ can still produce a small frequency shift.


\section{Qubit and coupler spectroscopy}
\label{app:spectro}

\begin{figure}
    \centering
    \includegraphics[width=\columnwidth]
    {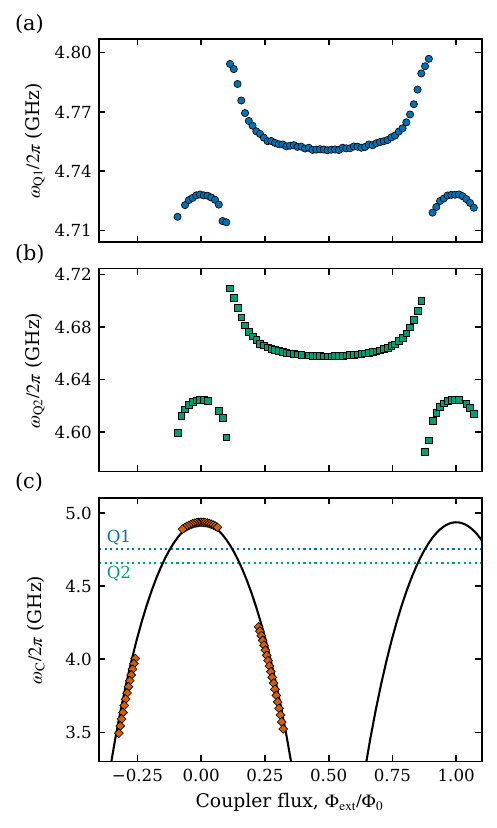}
\caption{Transition frequencies extracted from spectroscopy measurements as a function of the applied coupler flux $\Phi_{\mathrm{ext}}/\Phi_0$. 
(a) The transition frequency of Q$_1$.
(b) The transition frequency of Q$_2$. 
(c) The coupler transition frequency extracted from spectroscopy (orange markers) together with the fitted flux dependence (solid line). The dotted horizontal lines indicate the bare qubit frequencies $\omega_{Q1}/2\pi$ and $\omega_{Q2}/2\pi$.}
    \label{fig:qubit_spec}
\end{figure}

Figure~\ref{fig:qubit_spec} shows the measured qubit and coupler transition frequencies as a function of the external flux applied to the tunable coupler. Figures~\figpanelNoPrefix{fig:qubit_spec}{a} and \figpanelNoPrefix{fig:qubit_spec}{b} show the extracted transition frequencies of Q$_1$ and Q$_2$, respectively, while \figpanel{fig:qubit_spec}{c} shows the coupler transition frequency together with the fitted flux dependence used to map the applied flux to the corresponding coupler frequency.


\section{ZZ interaction}
\label{app:ZZ}

\begin{figure}
    \centering
    \includegraphics[width=\columnwidth]
    {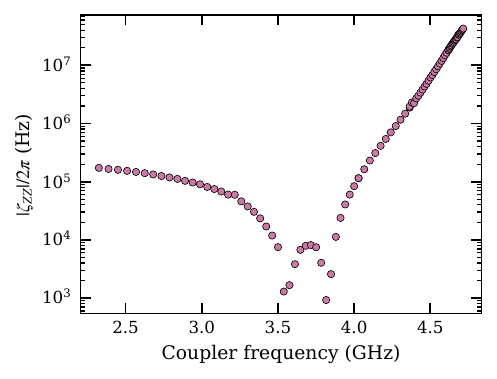}
\caption{Measured magnitude of the static $ZZ$ interaction, $|\zeta_{ZZ}|/2\pi$, as a function of coupler frequency.}
    \label{fig:ZZ}
\end{figure}

In addition to the microwave-drive crosstalk characterized through the cross-Rabi measurements, the device also exhibits a static $ZZ$ interaction between the two qubits~\cite{fors2024comprehensive, chu2021coupler}. We measure the ZZ interaction via the JAZZ protocol, an echo-based method for measuring the residual ZZ coupling~\cite{ku2020suppression, shirai2023all, li2024realization}. Figure~\ref{fig:ZZ} shows the measured magnitude of the $ZZ$ interaction as a function of coupler frequency. The $ZZ$ interaction can be strongly suppressed by an appropriate choice of coupler frequency, with one of the low-$ZZ$ operating points lying close to the frequency range of the measured crosstalk nulls of $R_{\mathrm{Q2}}^{\mathrm{cross}}$ and $R_{\mathrm{Q1}}^{\mathrm{cross}}$. Since the crosstalk-null frequencies can be shifted through the drive-to-coupler and drive-to-victim capacitances, the approach presented in this work provides a route to engineering the microwave crosstalk nulls toward a low-$ZZ$ operating point.


\section{Effective qubit-qubit coupling}
\label{app:Jeff}

\begin{figure}
    \centering
    \includegraphics[width=\columnwidth]{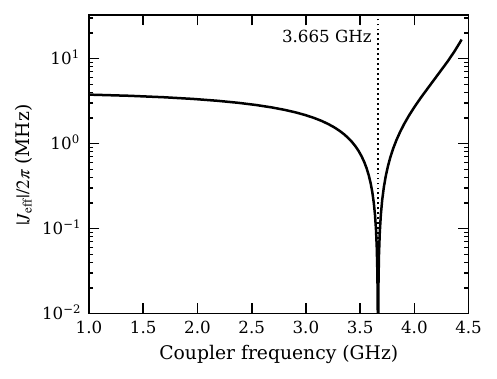}
    \caption{Effective qubit-qubit coupling $J_\text{eff}$ as a function of the coupler frequency. The dashed line indicates the coupler frequency for which this effective coupling is cancelled.}
    \label{fig:J_eff}
\end{figure}

Figure~\ref{fig:J_eff} shows the simulated effective coupling $J_\text{eff}$ for the two floating qubits in our processor. The model, described by Eq.~\ref{Jeff_eq}, is derived from the Schrieffer--Wolff transformation without taking into account either the drive or the parasitic couplings of the drive lines to the coupler and the victim qubit. The flux dependence of the qubit-coupler couplings is incorporated through $g_{i\text{C}}$, while the direct coupling $g_{12}$ remains constant. The plot demonstrates that there exists a coupler frequency (dashed line) for which this effective coupling is cancelled. 


\section{Comparing analytical results with numerical simulations for cross-Rabi frequencies}
\label{app:analytics_vs_numerics}

\begin{figure}
    \centering
\includegraphics[width=\columnwidth]{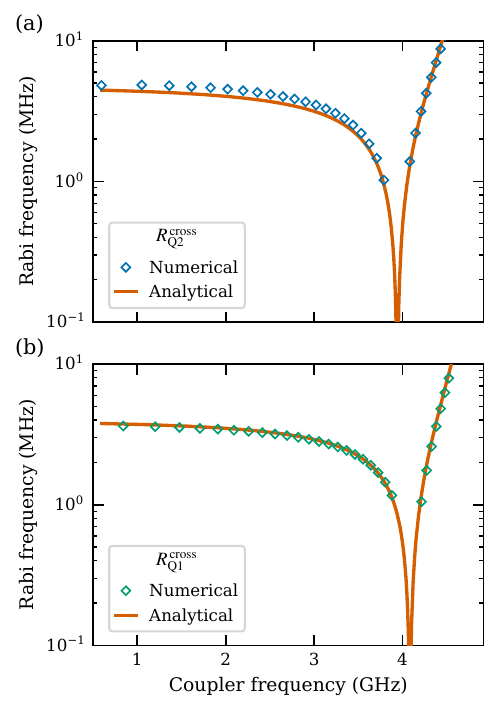}
\caption{%
Comparison of analytical and numerical cross-Rabi frequencies as a function of coupler frequency for 
(a) $R_{\mathrm{Q2}}^{\mathrm{cross}}$ and 
(b) $R_{\mathrm{Q1}}^{\mathrm{cross}}$.
Numerical results are shown by diamonds, while the analytical predictions are shown by solid lines.
}
\label{fig:analytics_vs_numerics}
\end{figure}

The analytical model for the cross-Rabi response developed in the main text, \secref{sec:model}, is obtained using a high-frequency expansion truncated at first order. To assess the accuracy of this truncation, we compare the analytical predictions with full numerical simulations. Figure~\ref{fig:analytics_vs_numerics} shows \(R_{\mathrm{Q2}}^{\mathrm{cross}}\) and \(R_{\mathrm{Q1}}^{\mathrm{cross}}\), calculated using the fitted parameters listed in \tabref{tab:drive_model_hierarchy}. The analytical and numerical results show very good agreement over the full coupler-frequency range. This comparison supports the validity of the first-order high-frequency expansion used in the analytical model. The numerical simulation procedure used for this comparison is described below.

We model the system of two qubits and a coupler using the full Hamiltonian in \eqref{eq:kerr_h}, retaining counter-rotating anharmonic and capacitive coupling terms. Each transmon is described using its four lowest oscillator states, \(\{|0\rangle,|1\rangle,|2\rangle,|3\rangle\}\), while the composite Hilbert space is truncated to a maximum of three total excitations. The microwave drive applied through a selected control line couples to both qubits and the coupler, with amplitudes determined by the drive capacitances and the inverse capacitance matrix. For each coupler flux, the undriven Hamiltonian is diagonalized and the dressed computational states are identified from their overlap with the corresponding bare states, using the Gale--Shapley algorithm~\cite{Gale1962}, as explained in Ref.~\cite{fors2024comprehensive}. 

Since the driven Hamiltonian is periodic with period \(T=2\pi/\Omega\), the dynamics are evaluated using Floquet theory~\cite{grifoni1998driven}, and the resulting state is projected onto the dressed computational basis to obtain the time-dependent populations and Rabi dynamics. The AC-Stark shift of the driven transition is accounted for numerically by adjusting the drive frequency to the minimum Floquet quasi-energy splitting, ensuring that the drives are resonant in each simulation.


\section{Comparing experimental and simulated Rabi dynamics}
\label{app:exp_vs_simulation_2D}

\begin{figure*}
    \centering
    \includegraphics[width=\textwidth]{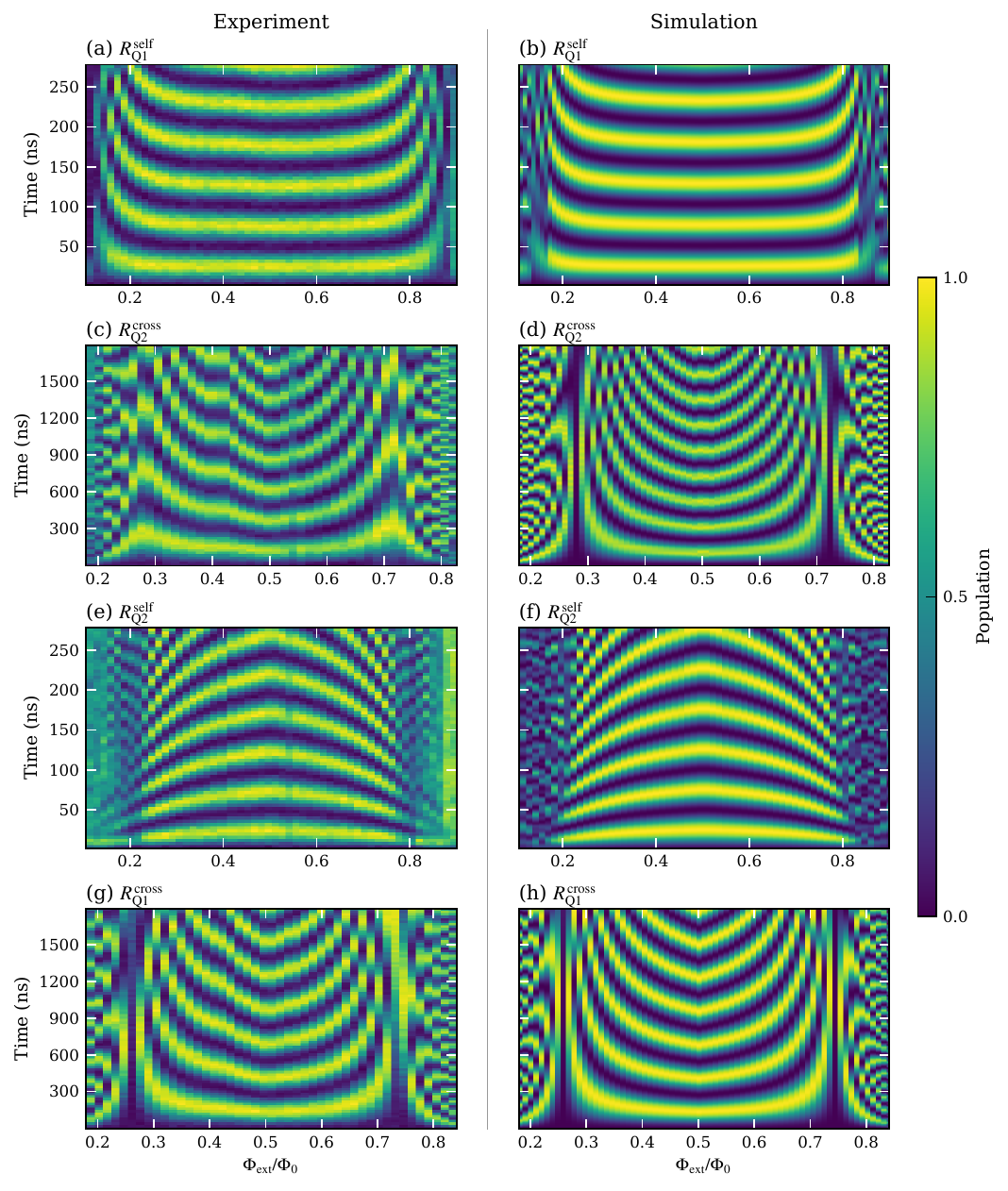}
    \caption{Measured (left column) and simulated (right column) Rabi oscillations of qubit populations (normalized in each row for clarity) as a function of coupler flux. For driving through XY$_1$, 
    (a, b) show driving of Q$_1$ through XY$_1$ (giving the Rabi frequency $R_{\rm Q1}^{\rm self}$) and 
    (c, d) show driving of Q$_2$ through XY$_1$ (giving the Rabi frequency $R_{\rm Q2}^{\rm cross}$). For driving through XY$_2$, 
    (e, f) show driving of Q$_2$ (giving the Rabi frequency $R_{\rm Q2}^{\rm self}$) and 
    (g, h) show driving of Q$_1$ (giving the Rabi frequency $R_{\rm Q1}^{\rm cross}$). 
    The simulations use the fitted parameters from \tabref{tab:drive_model_hierarchy}.}
    \label{fig:flux_vs_xtalk_2d}
\end{figure*}

In \figref{fig:flux_vs_xtalk_2d}, we compare the measured and simulated self- and cross-Rabi dynamics as a function of coupler flux. The plots in the left (right) column show the measured (simulated)  Rabi oscillations. The simulation results are obtained using the procedure detailed in \appref{app:analytics_vs_numerics} with the extracted parameters from \tabref{tab:drive_model_hierarchy}.

Overall, the simulated and measured Rabi oscillations show excellent agreement. A difference appears for $R_{\rm Q2}^{\rm cross}$ near its crosstalk-null frequency, where the simulation predicts stronger suppression than observed experimentally. The remaining finite crosstalk likely arises from additional microwave-crosstalk contributions not included in the present circuit model.


\clearpage

\bibliography{apssamp}

\end{document}